\documentclass{aa}  

\usepackage{graphicx}
\usepackage{placeins}

\usepackage{txfonts}
\usepackage{xcolor}

\usepackage{float}
\usepackage{hyperref}
\hypersetup{
    colorlinks=true,
    linkcolor=blue,
    filecolor=magenta,
    citecolor=blue,
    urlcolor=blue,
    pdftitle={The inner dark-matter structure of TNG50 simulated galaxies}}

\begin{document}

   \title{The inner dark-matter structure of TNG50 simulated galaxies}

   \author{Vicente Honorato\inst{1,2}\fnmsep\thanks{vicente.honorato@usm.cl}
          \and
          Antonio D. Montero-Dorta\inst{3}
          \and
          M. Celeste Artale\inst{4}
          \and
          Ankit Kumar\inst{4}
          }

   \institute{Departamento de Física, Universidad Técnica Federico Santa María, Casilla 110-V, Avda. España 1680, Valparaíso, Chile.
        \and
    Instituto de Física, Pontificia Universidad Católica de Valparaíso, Casilla 4950, Valparaíso, Chile.
        \and
    Departamento de Física, Universidad Técnica Federico Santa María, Avenida Vicuña Mackenna 3939, San Joaquín, Santiago, Chile.
         \and
    Universidad Andres Bello, Facultad de Ciencias Exactas, Departamento de Fisica y Astronomia, Instituto de Astrofisica, Fernandez Concha 700, Las Condes, Santiago RM, Chile.
             }

   \date{Received 5 April 2026 / Accepted 31 August 2026}

  \abstract 
    {In the framework of the $\Lambda$ cold dark matter model, galaxies evolve within dark matter (DM) haloes, where baryonic processes can modify the inner structure of the DM distribution. In particular, the condensation of baryons and associated feedback mechanisms can alter the inner density profiles of haloes, motivating detailed studies of their central regions.}  
   {We investigate the inner slope of the DM density profiles of galaxies in the TNG50 simulation, exploring its connection with galaxy properties and its evolution across cosmic time. We also assess the impact of baryonic processes on the inner DM structure by comparing galaxies to their counterparts in a DM-only (DMO) realisation of the same simulation.}
   {Spherically averaged DM density profiles were constructed for galaxies in the TNG50 simulation and their counterparts in the corresponding DMO run. The inner slope was quantified using an inner linear fit, defined as a direct power-law fit to the central region of the density profiles and motivated by the asymptotic behaviour of generalised Navarro--Frenk--White models. Subhaloes were matched between simulations and tracked across redshifts $z=0$, $0.2$, $0.7$, and $1$.}
   {The inner DM structure of galaxies in TNG50 shows that high-stellar-mass systems ($M_\star \gtrsim 10^{11}$ M$_\odot$) exhibit shallow inner slopes regardless of whether they are centrals or satellites, while lower-mass galaxies ($M_\star \lesssim 10^{9}$ M$_\odot$) show a broader diversity of profiles. At fixed stellar mass, low-mass satellites tend to be more cuspy, with the steepest slopes found in redder systems with lower $V_{\rm max}$ in more massive host haloes. At the population level, we find a modest cosmic evolution, with steeper profiles towards low redshifts in the hydrodynamical run and the opposite trend in DMO, while hydrodynamical galaxies remain systematically steeper at all redshifts. Finally, we verified that the fiducial interval provides the most suitable reference for measuring the shallowest inner slopes of DM density profiles at the TNG50 resolution limit.}
   {}


   \maketitle

\section{Introduction}

According to the standard $\Lambda$ cold dark matter (CDM) model, galaxies form at the centres of dark matter (DM) haloes as baryons cool and condense within the gravitational potential wells generated by the underlying DM distribution \citep{1978MNRAS.183..341W, 1991ApJ...379...52W}. Within this framework, \citet{1999ApJ...524L..19M} introduced the concept of `haloes within haloes'.\ They correspond to substructures known as subhaloes, which host galaxies embedded within larger systems. Subsequent studies using N-body simulations have shown that these systems naturally assemble into groups and clusters, where a central host halo is surrounded by a population of satellite subhaloes in orbit \citep{Diemand_2007, 2008MNRAS.391.1685S, 2019Galax...7...81Z}.

Within these haloes, the DM component follows characteristic radial density distributions that can be described by a variety of parametric profiles (e.g. \citealt{1965TrAlm...5...87E, 1987gady.book.....B, 1995ApJ...447L..25B, 1997ApJ...490..493N}). Using N-body simulations, \citet{1997ApJ...490..493N} showed that DM haloes are well described by a universal density profile featuring a cuspy inner region and a steeper outer slope. The Navarro--Frenk--White (NFW) profile is characterised by a smooth transition between these two regimes at a characteristic scale radius and provides a good description of haloes across a wide range of masses and redshifts. These results suggest that the internal structure of DM haloes emerges from a largely universal formation process driven by hierarchical clustering, highlighting the predictive power of N-body simulations in reproducing the large-scale structure of the Universe.

Baryons, unlike DM, dissipate energy through radiative cooling, which allows them to condense towards the centres of galaxies and within DM haloes \citep{1986ApJ...301...27B}. The gravitational interaction between baryons and DM induces a back-reaction that can modify the underlying halo potential, leading to changes in the central DM distribution. A number of studies have shown that baryonic processes associated with galaxy formation can have a non-negligible impact on the internal structure of DM haloes \citep[see e.g.][]{2009MNRAS.395L..57P, 2010MNRAS.405.2161D, 2010MNRAS.407..435A, 2019A&A...622A.197A}. Capturing these effects therefore requires numerical models that self-consistently follow the coupled evolution of baryons and DM.

The inclusion of baryonic physics in numerical simulations has led to substantial advances in modelling the large-scale Universe, allowing deviations from the predictions of pure N-body simulations to be explored in a self-consistent manner \citep{2020MNRAS.495.4800A}. In particular, hydrodynamical simulations capture the gravitational coupling between baryons and DM, showing that the condensation of baryons and subsequent feedback processes can significantly modify the DM distribution within haloes, often leading to more concentrated central profiles \citep{2004ApJ...616...16G, 2015MNRAS.452..343S, 2023MNRAS.520.2867V, 2025JCAP...07..026V}. These developments highlight the necessity of accounting for baryonic effects in order to establish physically meaningful and predictive models of galaxy formation and evolution.

The inner DM regions of galaxies remain an active area of research due to the long-standing core--cusp dichotomy, whose interpretation depends on both the physical processes involved and the methods used to characterise the inner density distribution \citep{2010AdAst2010E...5D}. On the one hand, baryonic feedback processes, such as supernova-driven outflows, have been shown to redistribute DM and transform initially cuspy profiles into shallower cores \citep{2012MNRAS.421.3464P}. On the other hand, alternative scenarios, including self-interacting DM, provide additional mechanisms to account for the observed diversity of inner halo structures \citep{2015MNRAS.453...29E}. In this context, recent numerical studies have increasingly adopted the inner density slope as a quantitative diagnostic to characterise the core--cusp dichotomy and to trace the response of DM to baryonic processes. High-resolution hydrodynamical simulations show that variations in the inner slope are closely linked to the balance between baryon condensation and feedback-driven gas removal: efficient star formation favours contraction and steeper slopes, while bursty feedback episodes can drive a progressive flattening of the inner profile \citep{2014MNRAS.437..415D, 2016MNRAS.456.3542T, 2020MNRAS.497.2393L}. However, at sufficiently low stellar masses, the limited energy available from stellar feedback may be insufficient to significantly modify the central DM distribution \citep{2012ApJ...759L..42P}. At the same time, recent work has highlighted that the inner DM density structure of galaxies exhibits a substantial halo-to-halo diversity, even among systems with intrinsically cuspy profiles, reflecting differences in their assembly history and dynamical state \citep{2018MNRAS.474.1398G}. More generally, numerical analyses have shown that these properties are sensitive to the interplay between baryonic processes and the underlying DM physics, giving rise to a wide range of inner density slopes \citep{2026A&A...708A..47D}.

Observational constraints from dwarf galaxies to galaxy clusters do not yet provide a unified picture of the inner structure of DM haloes. In the low-mass regime, H\,{\sc i} rotation curves of nearby dwarf galaxies and stellar-dynamical analyses of classical and ultra-faint dwarf spheroidals generally favour shallow or cored inner profiles over NFW-like cusps \citep{2008ApJ...681L..13B,2011ApJ...742...20W,2015AJ....149..180O,2024ApJ...973L..15S,2025A&A...699A.347A}. Other studies point to a broader diversity. Non-spherical dynamical models favour cuspy profiles in most classical dwarf spheroidals \citep{2020ApJ...904...45H}, while recent analyses of Draco infer cusp-like slopes with best-fitting values below but close to unity and statistically consistent with the NFW expectation \citep{2024ApJ...970....1V,2025ApJ...993..249Y}. The diversity of the inferred profiles may partly reflect degeneracies and differences in dynamical modelling \citep[e.g.][]{2011MNRAS.411.2118A}. Physical differences associated with their star-formation histories may also contribute, since dwarfs that ceased forming stars early are inferred to have central densities consistent with CDM cusps \citep{2019MNRAS.484.1401R}.

At higher mass scales, joint strong-lensing and stellar-dynamical modelling of massive early-type galaxies reveals a broad range of inner DM profiles, with most systems favouring steep slopes but a smaller subset consistent with much shallower distributions \citep{2018MNRAS.476..133O}. At cluster scales, analogous analyses that combine strong gravitational lensing, including constraints from radial arcs, with stellar kinematics of the brightest cluster galaxy have generally inferred inner DM profiles shallower than the NFW expectation \citep{2004ApJ...604...88S,2008ApJ...674..711S,2013ApJ...765...25N,2025MNRAS.541.2341C}.

In this context, the inner slope of DM density profiles provides a sensitive probe of the interplay between baryonic processes and environmental effects. This is particularly relevant for galaxies, where tidal interactions and gas depletion associated with the host halo can modulate the impact of baryonic physics on the central DM distribution. In this work, we investigated the inner regions of the DM density profiles of galaxies -- and, consequently, subhaloes -- in the IllustrisTNG\footnote{\url{http://www.tng-project.org}} hydrodynamical simulation. We propose a simple yet accurate method for characterising these internal profiles and evaluating the effect of baryons on them. We also analysed the dependence of the inner profiles on galaxy properties and their redshift evolution.

This paper is organised as follows. Section~\ref{sec:data} describes the IllustrisTNG simulation used in this work, the selection of the subhalo sample, the construction of the DM density profiles, and the matching between the hydrodynamical and DM-only (DMO) runs. In Sect.~\ref{subsec:models}, we introduce the density profile models used to characterise the inner structure of the DM component in galaxies. Section~\ref{sec:results} presents the main results of our analysis, focusing on the dependence of the inner density slopes on galaxy properties, environment, and redshift. Finally, Sect.~\ref{sec:discussion} discusses the physical interpretation of our findings, and Sect.~\ref{sec:conclusions} summarises the main conclusions.

\section{Data} \label{sec:data}

\subsection{TNG50 simulation} \label{subsec:TNG50_1_sim}

This study is based on the TNG50 simulation, the highest-resolution magnetohydrodynamical run of the IllustrisTNG suite \citep{2019MNRAS.490.3234N,2019MNRAS.490.3196P}. The simulation adopts a $\Lambda$CDM cosmology consistent with the \citet{2016A&A...594A..13P} results, with parameters $\Omega_{m,0}=0.3089$, $\Omega_{b,0}=0.0486$, $\Omega_{\Lambda}=0.6911$, $H_{0}=100\,h\,\mathrm{km\,s^{-1}\,Mpc^{-1}}$ with $h=0.6774$, $\sigma_{8}=0.8159$, and $n_{s}=0.9667$. TNG50 follows the evolution of cosmic structure from high redshifts to $z=0$ within a comoving volume of $(35\,h^{-1}\,\mathrm{cMpc})^{3}$, achieving a DM particle mass resolution of $3.1\times10^{5}\,\mathrm{M}_\odot\,h^{-1}$ and a typical baryonic mass resolution of $\sim 5.7\times10^{4}\,\mathrm{M}_\odot\,h^{-1}$. The high mass and spatial resolution of TNG50 enable the internal structure of individual galaxies and their subhaloes to be resolved in detail, including the central regions relevant for studies of inner density profiles. The simulation adopts a Plummer-equivalent gravitational softening length of $\epsilon_{\rm DM}=0.29$ physical kpc for DM particles throughout the redshift range considered in this work \citep{2019MNRAS.490.3196P}.

The IllustrisTNG project was carried out using the moving-mesh code \textsc{AREPO} \citep{2010MNRAS.401..791S}, which solves the equations of magnetohydrodynamics with a finite-volume scheme on a dynamic, unstructured Voronoi tessellation. This approach provides adaptive spatial resolution and an accurate treatment of complex gas dynamics. The simulations include a comprehensive set of baryonic physics models, such as radiative cooling and heating, star formation, stellar feedback, and supermassive black hole growth and feedback, implemented through sub-grid prescriptions calibrated to reproduce a wide range of observed galaxy properties and statistical trends \citep{2017MNRAS.465.3291W,2018MNRAS.473.4077P}. The DM haloes are first identified using a friends-of-friends (FoF) algorithm, while gravitationally bound substructures within each FoF group are subsequently identified with the \textsc{SUBFIND} algorithm. In this framework, each FoF halo hosts a central galaxy associated with the most massive subhalo, whereas the remaining subhaloes correspond to satellite galaxies orbiting within the host halo.

This balance between resolution and cosmological volume is particularly well suited to the aims of this work, providing the spatial detail required to probe the inner structure of galaxies, including low-mass systems, while retaining a broad and consistently selected sample for investigating population-level trends across masses, redshifts, and environments. In this respect, TNG50 provides an intermediate approach between larger-volume cosmological runs and high-resolution zoom-in suites.

\subsection{Subhalo sample} \label{subsec:subhalo_sample}

From TNG50, we constructed the galaxy sample analysed in this work by selecting gravitationally bound subhaloes\footnote{Throughout this work, the term `subhalo' is used to refer to both central and satellite galaxies, without distinction.} and applying selection criteria aimed at ensuring that the relevant physical properties are not affected by numerical resolution limitations. In particular, we required galaxies at $z=0$ to have stellar masses corresponding to at least 50 initial gas cells, which sets a lower stellar-mass limit of $4.25\times10^{6}\,\mathrm{M}_\odot$. In addition, we imposed a resolution criterion on the total galaxy mass by requiring each system to contain at least 50 DM particles, corresponding to a minimum galaxy mass of $2.25\times10^{7}\,\mathrm{M}_\odot$. These thresholds are consistent with previous studies \citep{2019MNRAS.489.2634H,2020MNRAS.496.1182M,2021MNRAS.508..940M} and define a galaxy sample that is well resolved in both its stellar and DM components.

For each selected galaxy, we computed the DM density profile using the particles gravitationally bound to its associated subhalo in the TNG50 simulation. The centre of each system was defined as the position of the particle with the minimum gravitational potential, accounting for the periodic boundary conditions of the simulation volume. The DM particles were then binned in concentric spherical shells to estimate the density as a function of radius, using a logarithmic radial grid extending from $0.1\,\mathrm{kpc}$ to the maximum radius of each system. In addition to the density profiles, we extracted several global properties from the simulation catalogue that were used in the analysis presented in Sect.~\ref{sec:results}. These include the stellar mass ($M_\star$; in units of $\mathrm{M}_\odot$), defined as the total mass of stellar particles gravitationally bound to the subhalo; the host-halo mass ($M_{\rm host}$; in $\mathrm{M}_\odot$), which corresponds to the virial mass of the FoF group to which the galaxy belongs; the maximum circular velocity ($V_{\rm max}$; in $\rm km \, s^{-1}$); the specific star formation rate (sSFR; in $\rm yr^{-1}$); and the $(g-r)$ rest-frame colour computed from synthetic photometry.

Finally, in order to study the redshift evolution of the galaxy population selected at $z=0$, we reconstructed the evolutionary history of each system using the corresponding \textsc{LHaloTree} merger trees in the simulation (e.g. \citealt{2015MNRAS.449...49R,2019ComAC...6....2N}) by starting from its \texttt{SubhaloID} in the final snapshot and following the main progenitor branch backwards in time via the \texttt{FirstProgenitorID} pointer at each snapshot. This uniquely associated each present-day galaxy with its progenitors at higher redshifts, where all structural and global properties were consistently measured. To assess the impact of baryonic physics on galaxy structure, each system was additionally matched to its counterpart in the corresponding DMO run using the precomputed Hydro--DMO matching indices provided in the simulation data products (specifically, the \texttt{SubhaloIndexDark\_LHaloTree} field), which are constructed using the LHaloTree algorithm based on the overlap of DM particle IDs between the baryonic and DMO runs. Only bidirectional matches were accepted \citep[e.g.][]{2015A&C....13...12N}.

\begin{figure*}[h!]
    \sidecaption
    \includegraphics[width=12cm]{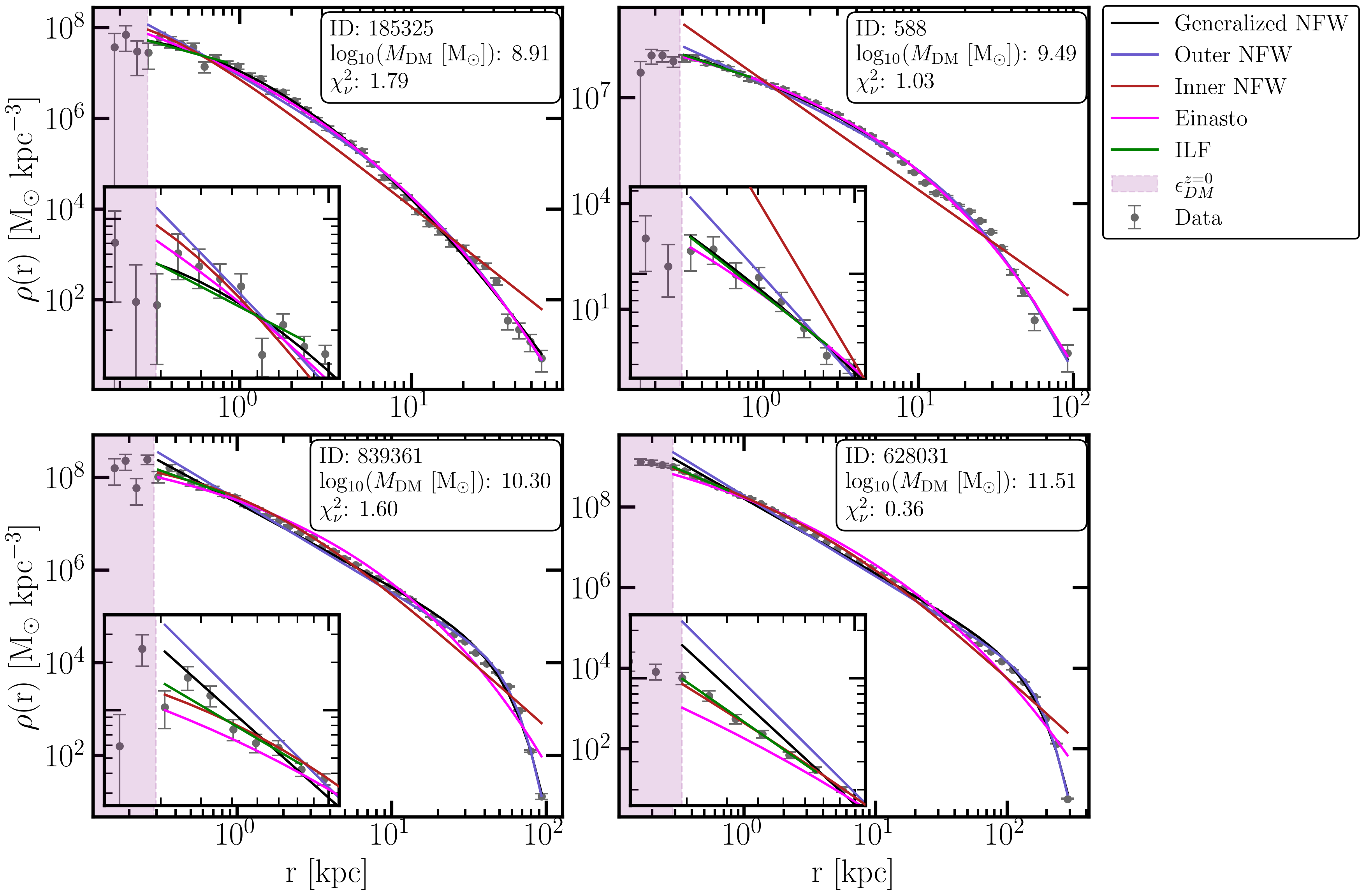}
    \caption{Examples of DM density profiles from our catalogue for galaxies spanning a range of DM masses, together with the five fitting models adopted in this work. \textit{Top panels}: Satellite galaxies. \textit{Bottom panels}: Central galaxies. Each panel shows the full radial profile, with an inset highlighting the inner region above the resolution limit. The shaded vertical region indicates radii below the adopted Plummer-equivalent gravitational softening length. Grey points with error bars represent the measured density profile, with uncertainties reflecting Poisson noise in the particle counts; the coloured curves show the corresponding best-fitting models. The ID reported in each panel corresponds to the \texttt{SubhaloID} of the galaxy in TNG50, and the reduced chi-square value ($\chi^2_\nu$) is obtained from the ILF.}
    \label{fig1:fits_all_prof}
\end{figure*}

\section{Density profile models} \label{subsec:models}

As an initial reference, we considered the Einasto profile \citep{1965TrAlm...5...87E}, which has become a standard description of simulated and observed DM haloes \citep[e.g.][]{2006AJ....132.2685M, 2010MNRAS.402...21N, 2014MNRAS.441.3359D, 2022A&A...667A..47B}. Its smooth curvature and flexible shape allow it to capture the global behaviour of many profiles without imposing fixed asymptotic slopes. The Einasto model is given by
\begin{align}
    \rho_{\rm Einasto}(r) = \rho_{-2}\,
    \exp\!\left[-2n\left(\left(\frac{r}{r_{-2}}\right)^{1/n}-1\right)\right],
\end{align}

\noindent where $\rho_{-2}$ is the density at the radius $r_{-2}$, and the parameter $n$ controls the radial variation of the logarithmic slope.

We also explored the generalised NFW (gNFW) profile (see \citealt{2010gfe..book.....M}),
\begin{align} \label{eq:gnfw}
\rho_{\rm gNFW}(r) =
\frac{\rho_0}{(r/r_s)^{\gamma}\,[1+(r/r_s)^{\alpha}]^{(\beta-\gamma)/\alpha}},
\end{align}

\noindent where $\gamma$, $\beta$, and $\alpha$ describe the inner slope, outer slope, and transition sharpness, respectively. This family includes the standard NFW profile for $(\alpha,\beta,\gamma)=(1,3,1)$ and encompasses commonly used variants. In particular, we adopted the `outer NFW' form, used in \citet{2024MNRAS.52711996H}, obtained by fixing the inner slope to $\gamma=2$,  
\begin{align}
    \rho_{\rm outerNFW}(r) =
    \frac{\rho_0}{(r/r_s)^2\,\left[1+(r/r_s)^{\alpha}\right]^{6}},
\end{align}

\noindent as well as the `inner NFW' variant \citep[e.g.][]{2019ApJ...887...94R,2024MNRAS.528..693O}, where only the inner slope is left free,
\begin{align}
\rho_{\rm innerNFW}(r) =
\frac{\rho_0}{(r/r_s)^{\gamma}\,(1+r/r_s)^{3-\gamma}}.
\end{align}

In addition to these parametric models, we considered an inner linear fit (ILF) to estimate the asymptotic inner slope of the density profile. The logarithmic slope of the gNFW density profile is given by
\begin{align}
\frac{d\ln\rho}{d\ln r}
= -\gamma - (\beta-\gamma) \frac{(r/r_s)^{\alpha}}{1+(r/r_s)^{\alpha}},
\end{align}

\noindent where $\gamma$ denotes the asymptotic inner slope. In the inner regime ($r \ll r_s$), the second term vanishes and the logarithmic slope converges to a constant value,
\begin{align}
\frac{d\ln\rho}{d\ln r} \xrightarrow{r \ll r_s} -\gamma,
\end{align}

\noindent which corresponds to a simple power-law density profile,
\begin{align}
\rho(r \ll r_s) \propto r^{-\gamma}.
\end{align}

Because we analysed the density profiles in logarithmic space, this asymptotic behaviour can be expressed as a linear relation,
\begin{align}
\log_{10}\rho(r \ll r_s) = -\gamma\,\log_{10} r + \log_{10}\rho_{0,\mathrm{lin}}.
\end{align}
We therefore treated the estimation of the inner slope ($\gamma$) through a linear least-squares fit in $\log_{10}\rho$--$\log_{10}r$ space as an additional fitting approach. The density profiles were binned in logarithmic radial intervals, and the fit was performed using the non-linear least-squares algorithm implemented in the \textsc{SciPy} Python library. The fit was restricted to radii between $\epsilon_{\rm DM}$ and $3\epsilon_{\rm DM}$, where $\epsilon_{\rm DM}$ denotes the Plummer-equivalent gravitational softening length of the DM particles introduced in Sect.~\ref{subsec:TNG50_1_sim}. We adopted this radial range to make full use of the spatial resolution of TNG50 and probe the innermost structure accessible in the simulation, while excluding radii below the nominal softening scale. Since convergence studies indicate that the convergence radius of simulated DM density profiles can lie beyond the nominal softening scale \citep{2003MNRAS.338...14P}, we examine how the inferred slopes depend on the adopted radial fitting interval in Sect.~\ref{subsec:diff_r_initial}.

Throughout this work, the inner slope ($\gamma$) refers to the value obtained from the ILF. Because this estimate is based on the adopted sub-kiloparsec radial range, the classification of density profiles as `cuspy' or `core-like' refers exclusively to the logarithmic slope measured over this interval and does not imply the presence of a constant-density core in the classical sense.

\section{Results}\label{sec:results}

\subsection{TNG50 density profile fitting} \label{subsec:Profiles_in_TNG}

Figure~\ref{fig1:fits_all_prof} shows representative examples of the density profiles in our catalogue together with the five fitting functions adopted in this work. For each galaxy, we display the full radial profile and a zoomed-in view of the innermost region, restricted to radii above the resolution limit. The models tested include the gNFW, outer NFW, and inner NFW variants (all based on the original NFW form; \citealt{1997ApJ...490..493N}), the Einasto profile \citep{1965TrAlm...5...87E}, and the ILF described in Sect.~\ref{subsec:models}. These examples illustrate the diversity of profile shapes across the sample and the corresponding differences in the quality of the fits. In particular, the models vary in their ability to reproduce the inner slope and the transition between the central and outer regions, with performance depending primarily on DM mass and the extent of the adopted radial range. These trends are clearly visible in the examples shown, which highlight the contrasting behaviours of the different fitting functions across the sample.

The representative galaxies shown in Fig.~\ref{fig1:fits_all_prof} span a wide range of DM masses, from $M_{\rm DM}\simeq10^{8.9}\,{\rm M_\odot}$ up to $M_{\rm DM}\simeq10^{11.5}\,{\rm M_\odot}$. Across this interval, the inner density profiles exhibit a broad variety of behaviours, ranging from nearly power-law cusps to profiles showing mild central flattening within the adopted inner region. In order to characterise this diversity in a homogeneous and physically motivated way, we relied on the ILF, which provides a local estimate of the inner density slope over the innermost scales considered in this analysis.

\begin{figure}[h!]
    \centering
    \includegraphics[width=0.9\linewidth]{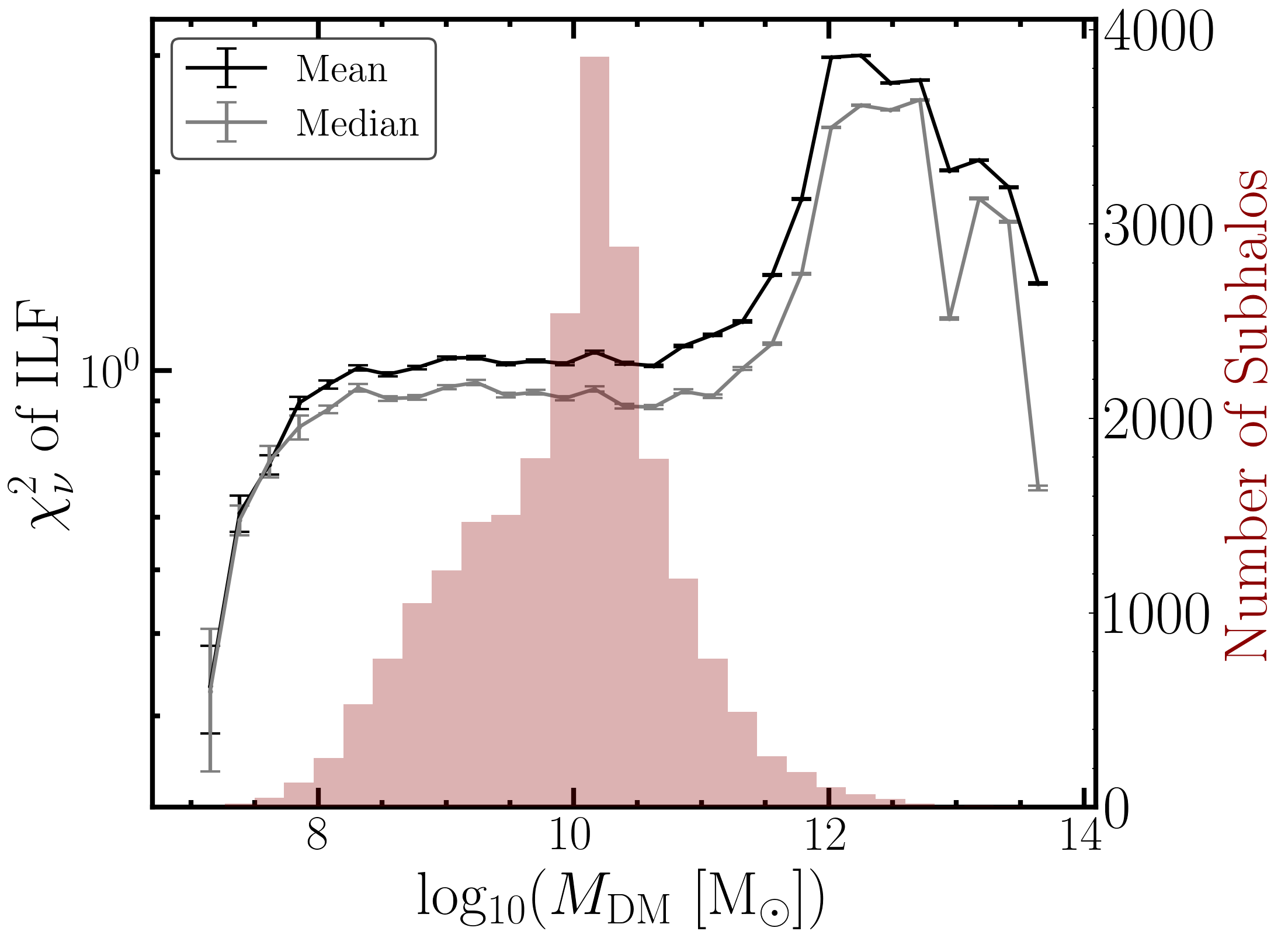}
    \caption{Reduced chi-square ($\chi^2_\nu$) of the linear fits as a function of galaxy DM mass. The black and grey curves show the mean and median in logarithmic mass bins, respectively, while the shaded histogram shows the underlying DM mass distribution of the sample. Error bars were estimated through bootstrap resampling.}
    \label{fig:chi2_mass}
\end{figure}

In what follows, we use the ILF as our reference model to estimate the inner slope ($\gamma$) of the density profiles. Unlike the parametric models considered above, which are anchored to the full radial extent of the profile and whose inner behaviour is therefore influenced by the transition to larger scales, the ILF provides a local estimate of the inner power-law slope over the adopted radial interval. By performing the fit locally in logarithmic space, this approach is less sensitive to the outer profile shape and provides a consistent characterisation of the innermost structure accessible within the resolution limits of the simulation.

The quality of these linear fits was quantified using the reduced chi-square ($\chi^2_\nu$), computed following the formulation adopted by \citet{2024MNRAS.52711996H}:

\begin{align}
\chi^2_\nu = \frac{1}{N-p}\sum_{i=1}^{N}
\left(
\frac{\ln\rho_{{\rm data},i}-\ln\rho_{{\rm fit},i}}
{\sigma_{\rho,i}/\rho_{{\rm data},i}}
\right)^2,
\end{align}

\noindent where $N$ is the number of radial bins, $p$ is the number of free parameters of the fit, and $\sigma_{\rho,i}$ denotes the uncertainty associated with each density measurement. The ratio $\sigma_{\rho,i}/\rho_{{\rm data},i}$ corresponds to the uncertainty propagated into natural-logarithmic space.

Each galaxy was assigned its own value of $\chi^2_\nu$ (as shown for representative cases in Fig.~\ref{fig1:fits_all_prof}), allowing us to examine how the goodness of fit varies across the mass range of the catalogue. Figure~\ref{fig:chi2_mass} displays the median and mean $\chi^2_\nu$ as a function of DM mass, together with the underlying mass distribution of the sample. The overall increase in $\chi^2_\nu$ with mass is consistent with the trend reported by \citet{2024MNRAS.52711996H}. The mass histogram illustrates the extended DM mass range covered by the present analysis, which aims to characterise an extended sample resolved in TNG50 using selection criteria motivated by the analysis of inner density profiles. Therefore, the low-mass regime was examined within the adopted sample selection.

\begin{figure}[h!]
    \centering
    \includegraphics[width=0.85\linewidth]{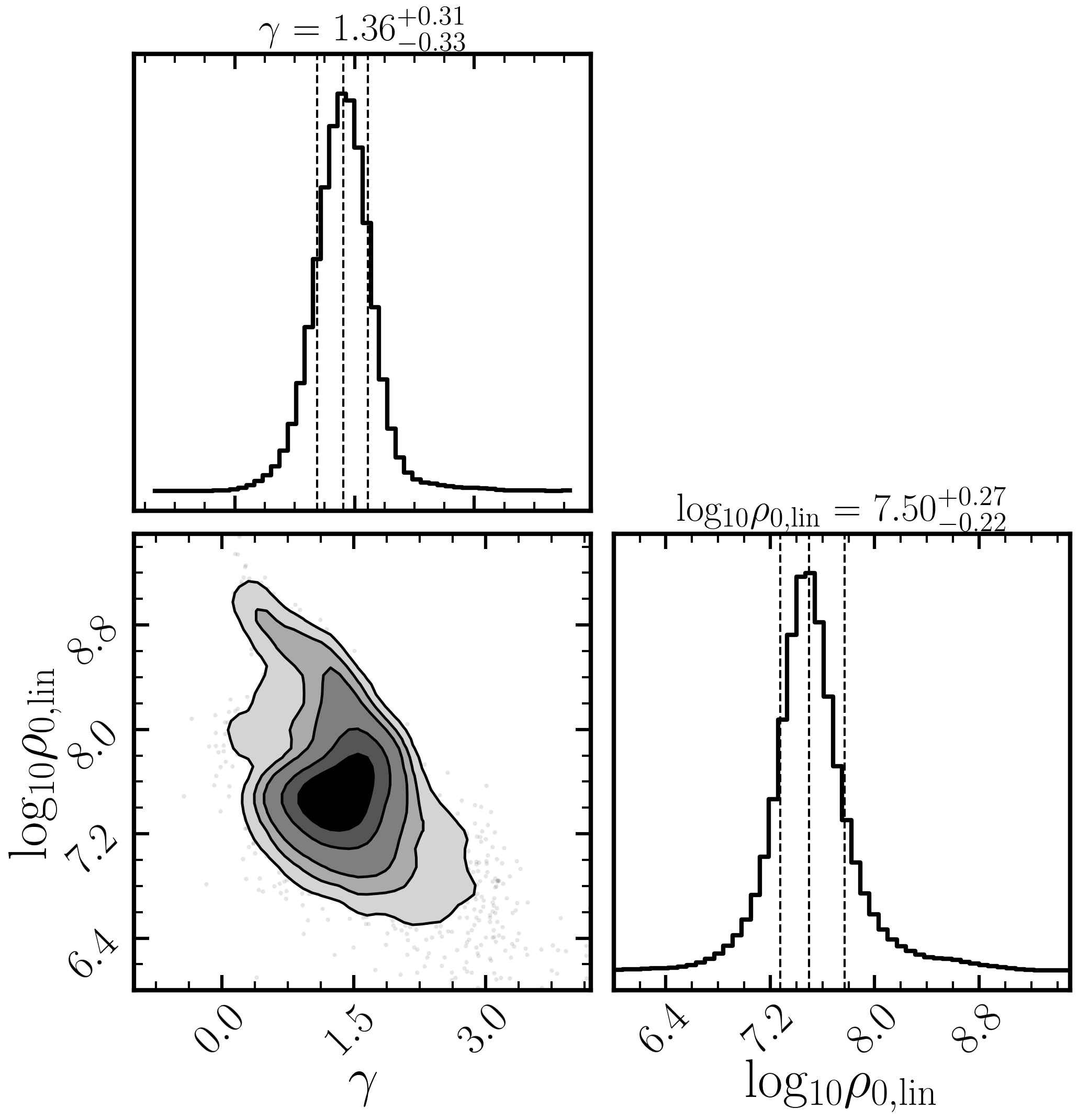}
    \caption{Distribution of the ILF parameters across the galaxy sample at $z=0$. \textit{Top and right}: Distributions of $\gamma$ and $\log_{10}\rho_{0,\mathrm{lin}}$. \textit{Bottom left}: Their joint distribution. Dashed lines mark the median and the 16th and 84th percentiles, and the contours enclose 50, 75, 90, 95, and 98.5\% of the smoothed distribution.}
    \label{fig:ilf_distribution}
\end{figure}

To characterise the distribution of the ILF parameters at $z=0$, Fig.~\ref{fig:ilf_distribution} shows the one-dimensional and joint distributions of $\gamma$ and $\log_{10}\rho_{0,\mathrm{lin}}$ across the galaxy sample. The distribution of inner slopes has a median of $\gamma=1.36$, with a 16th--84th percentile range of $1.03$--$1.67$, while the density normalisation has a median of $\log_{10}\rho_{0,\mathrm{lin}}=7.50$, with a corresponding range of $7.28$--$7.77$. The joint distribution suggests a mild anti-correlation, with steeper inner profiles tending to have lower density normalisations. This trend is more apparent in the outer contours, whereas the innermost contours remain comparatively compact and show only weak elongation. Collectively, the $\chi_\nu^2$ values close to unity over the mass range containing most of the sample and the well-defined distributions of $\gamma$ and $\log_{10}\rho_{0,\mathrm{lin}}$ support the use of the ILF to characterise the inner DM slopes over the adopted radial range.

\subsection{Dependence on galaxy properties} \label{subsec:connection_gal_properties}

\begin{figure}[h!]
    \centering
    \includegraphics[width=0.9\linewidth]{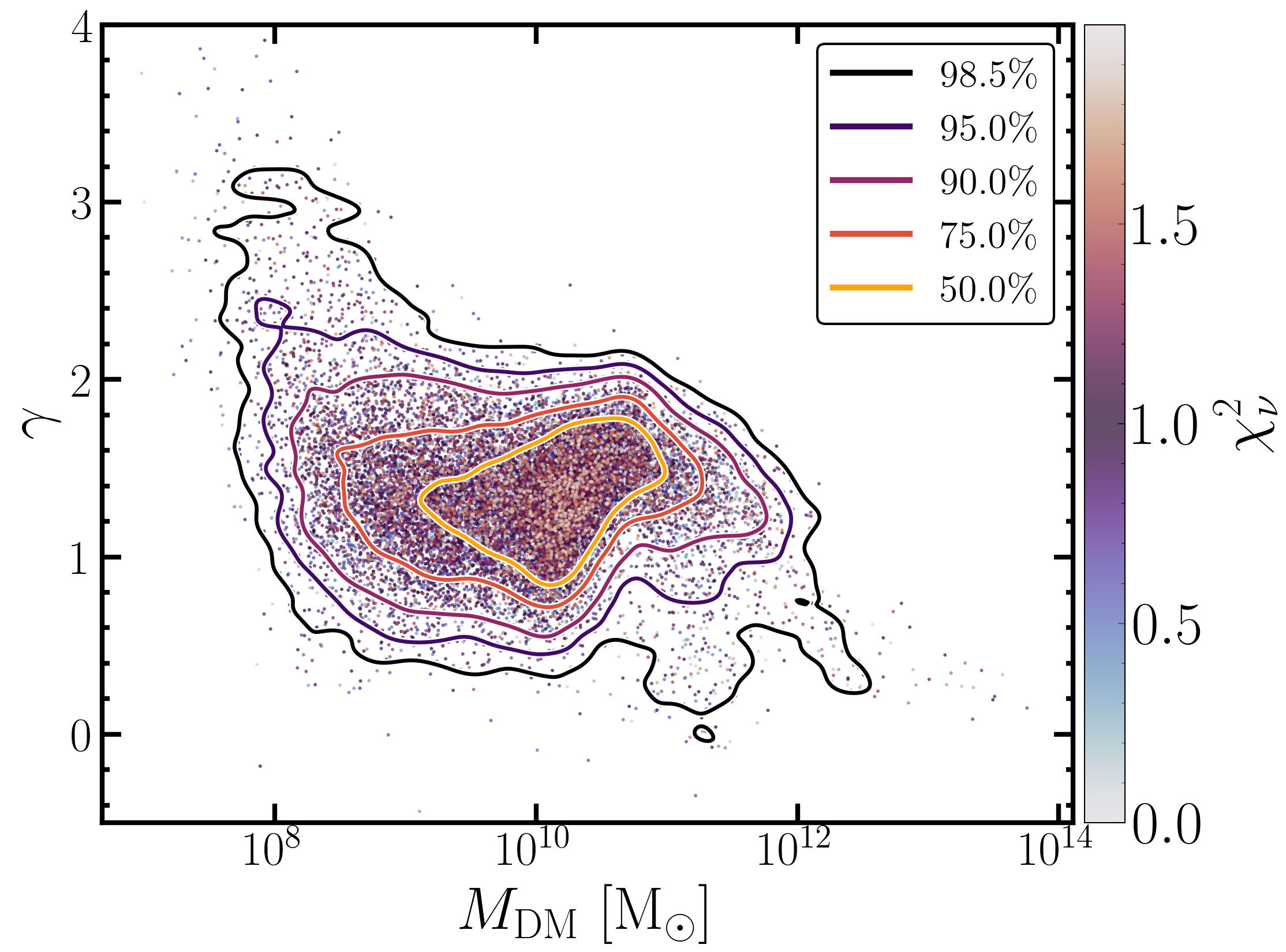}
    \caption{Two-dimensional distribution of the inner slope ($\gamma$) as a function of galaxy DM mass. Galaxies are coloured by the reduced chi-square ($\chi^2_\nu$) of the linear fits, while the contours show the kernel density estimate of the distribution.}
    \label{fig:kde}
\end{figure}

In order to explore physical trends in the inner DM structure, we examined how the inner slope ($\gamma$) varies with a range of galaxy and halo properties. Figure~\ref{fig:kde} presents the global relation between $\gamma$ and the galaxy DM mass. The distribution does not follow a simple monotonic trend. At the low-mass end ($M_{\rm DM}\lesssim10^{9}\,{\rm M_\odot}$), galaxies preferentially exhibit steep inner profiles, with typical values $\gamma \gtrsim 1.5$. In contrast, the most massive systems ($M_{\rm DM}\gtrsim10^{11}\,{\rm M_\odot}$) preferentially exhibit lower values of $\gamma$ than low-mass systems, typically $\gamma \lesssim 1.2$.

\begin{figure*}[h!]
    \sidecaption
    \includegraphics[width=12cm]{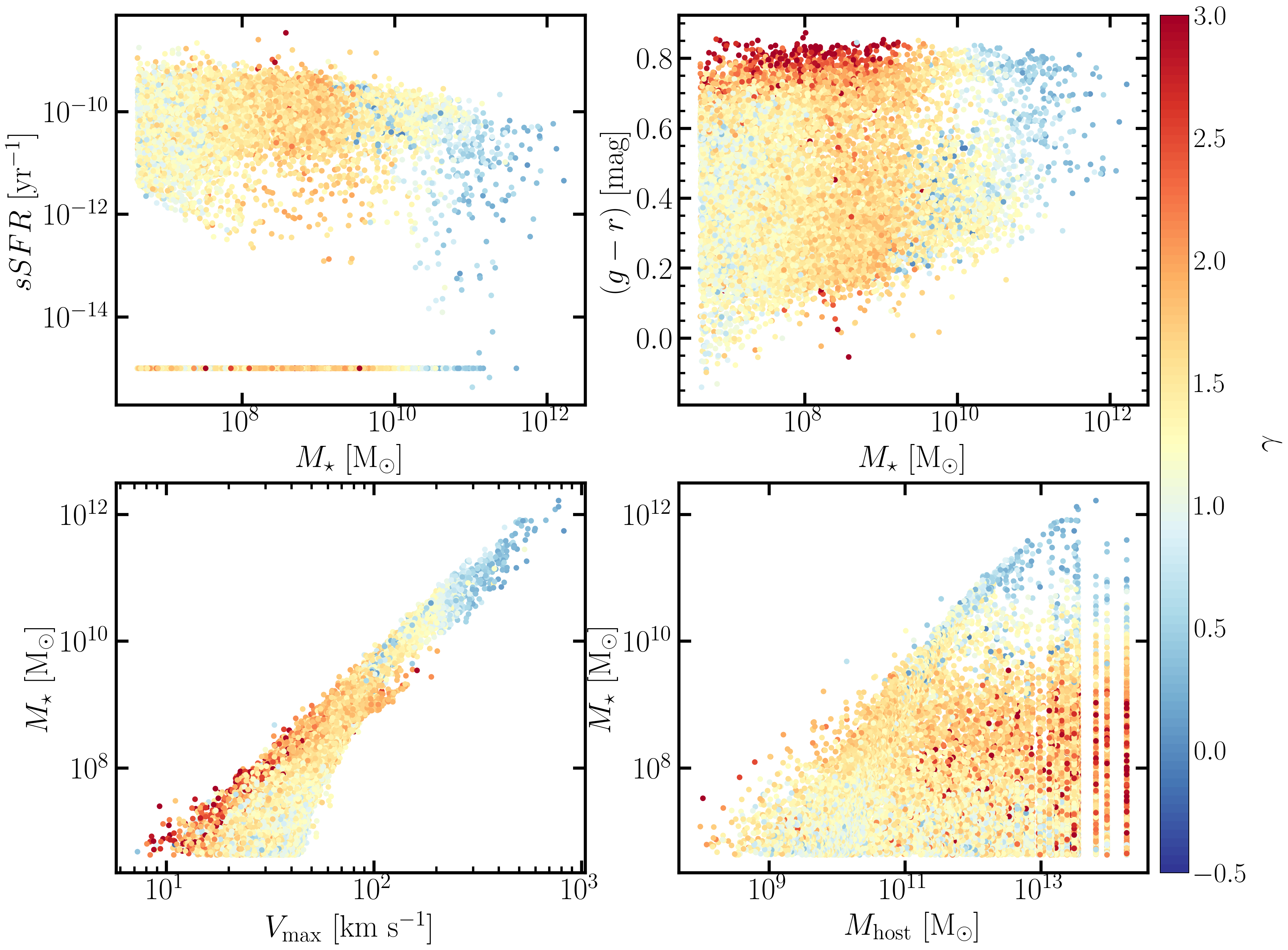}
    \caption{Galaxy properties colour-coded by the inner slope, $\gamma$. \textit{Top left}: sSFR versus stellar mass. \textit{Top right}: $(g-r)$ colour versus stellar mass. \textit{Bottom left}: Stellar mass versus maximum circular velocity, $V_{\rm max}$. \textit{Bottom right}: Stellar mass versus host-halo mass, $M_{\rm host}$.}
    \label{fig:2x2_properties}
\end{figure*}

Between these regimes, an extended intermediate-mass range ($M_{\rm DM}\sim10^{9}$--$10^{11}\,{\rm M_\odot}$) displays an approximately flat trend. In this interval, the contours of the kernel density estimate enclosing the highest point densities indicate that the bulk of the galaxy population (roughly the central $50$--$75\%$) is concentrated within $\gamma\simeq1$--$1.5$, with a mild tendency towards steeper slopes at higher masses. While the outer, lower-density contours reveal a broader dispersion in $\gamma$, they do not show a clear preference for either cuspy or core-like profiles. The colour-coding by the reduced chi-square ($\chi^2_\nu$) indicates that the linear fits are statistically acceptable across most of the mass range probed, with typical $\chi^2_\nu$ values of order unity.

To understand how this distribution maps onto galaxy observables, Fig.~\ref{fig:2x2_properties} summarises several key properties colour-coded by their corresponding inner slopes ($\gamma$). These diagrams provide a global view of how the inner DM structure is distributed across the sSFR--$M_\star$ plane \citep[e.g.][]{2007ApJ...660L..43N,2012ApJ...754L..29W,2014ApJS..214...15S}, the galaxy colour--$M_\star$ relation \citep[e.g.][]{2015MNRAS.452.2879T,2018MNRAS.475..624N}, the $M_\star$--$V_{\rm max}$ connection \citep[e.g.][]{2017MNRAS.464.2419S,2017MNRAS.464.4736F,2019MNRAS.490.3196P}, and the stellar-to-halo mass relation \citep[e.g.][]{2010ApJ...710..903M,2013MNRAS.428.3121M,2018AstL...44....8K,2020MNRAS.499.5656R}. The colour bar spans the range $-0.5 \leq \gamma \leq 3$, where values approaching $\gamma \simeq 0$ correspond to strongly flattened inner density profiles, while $\gamma \gtrsim 2$ indicates very steep, cusp-like inner slopes. Although the upper end of this range extends beyond values commonly quoted for asymptotic inner slopes in the literature, it reflects effective slopes measured over a finite radial interval. In this sense, the adoption of a simple power-law fit restricted to the innermost resolved region provides a sensitive probe of steep inner gradients that may be partially smoothed out when fitting full parametric profiles over broader radial ranges, and therefore allows baryonic processes such as adiabatic contraction to be captured more directly. Our fiducial choice of fitting from the gravitational softening scale balances resolution considerations with the goal of retaining genuinely inner structural information, while the impact of adopting more conservative inner radii is explicitly quantified in Sect.~\ref{subsec:diff_r_initial}. Across these panels, $\gamma$ exhibits coherent patterns, with low-mass, red, or low-$V_{\rm max}$ systems tending to display higher values of $\gamma$, whereas more massive or actively star-forming galaxies preferentially host shallower inner slopes. This motivates a more detailed inspection of the central and satellite populations separately.

\begin{figure*}[h!]
    \centering
    \includegraphics[width=0.9\linewidth]{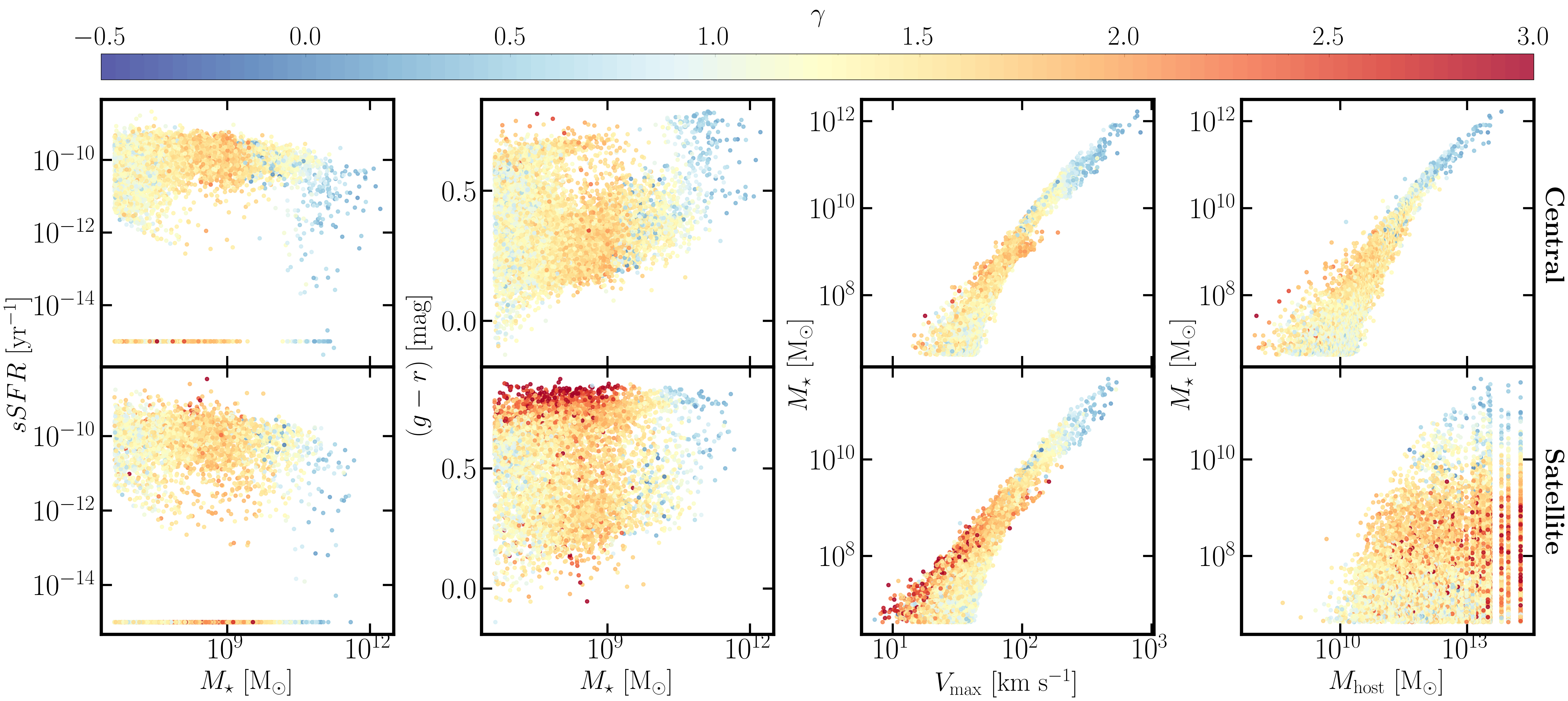}
    \caption{Scaling relations between galaxy and halo properties for central (\textit{top panels}) and satellite (\textit{bottom panels}) galaxies. \textit{Left to right}: sSFR versus stellar mass ($M_\star$); rest-frame colour $(g-r)$ versus $M_\star$; stellar mass versus maximum circular velocity ($V_{\rm max}$); and stellar mass versus host-halo mass ($M_{\rm host}$). Points are colour-coded by the inner density slope ($\gamma$) obtained from the linear fit.}
    \label{fig:2x4_figure}
\end{figure*}

Figure~\ref{fig:2x4_figure} presents a set of scaling relations between galaxy and halo properties, explicitly separating central (top panels) and satellite (bottom panels) galaxies. Each column corresponds to a different relation, while the colour-coding indicates the inner density slope ($\gamma$) derived from the linear fit to the central DM density profiles. This layout allows a direct comparison of how galaxy properties and inner halo structure differ between central and satellite systems, providing a first-order proxy for environmental and evolutionary effects.

\begin{figure}[h!]
    \centering
    \includegraphics[width=0.9\linewidth]{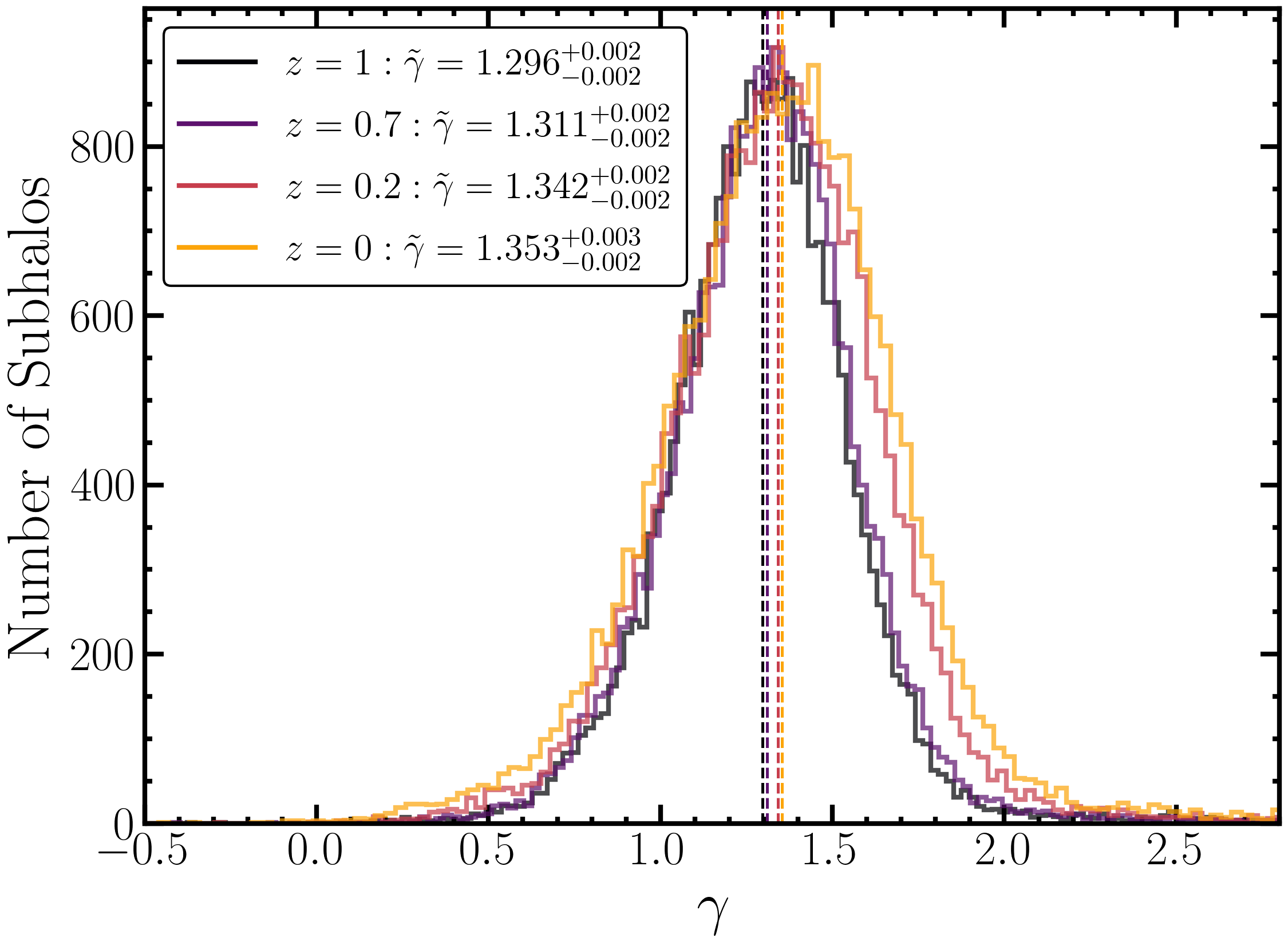}
    \caption{Evolution of the distribution of the inner slope ($\gamma$) from $z=1$ to $z=0$. Coloured histograms show the distribution of galaxies at each redshift. Dashed vertical lines mark the corresponding medians, with uncertainties indicating the 16th--84th percentile confidence intervals estimated via bootstrap resampling.}
    \label{fig:gamma_distribution_evol}
\end{figure}

The leftmost column shows the relation between the sSFR and stellar mass. Central galaxies populate the expected star-forming sequence, with a clear trend of decreasing sSFR towards higher stellar masses. Satellites, in contrast, display a slightly broader distribution, including a significant population of systems with suppressed star formation at fixed $M_\star$. In particular, many satellites occupy the quenched regime commonly defined by $\mathrm{sSFR}\lesssim10^{-11}\,\mathrm{yr}^{-1}$ \citep[e.g.][]{2004MNRAS.351.1151B,2019MNRAS.485.4817D}. Galaxies with numerically zero star formation appear as a horizontal sequence at the minimum sSFR value, corresponding to systems with strictly no ongoing star formation. No strong dependence of the inner slope on sSFR at fixed stellar mass is apparent for the central or satellite populations. Rather, variations in $\gamma$ are primarily driven by stellar mass, with higher-mass systems tending to exhibit shallower inner density slopes, largely independent of their current star-formation activity. This behaviour is common to both centrals and satellites.

The second column presents the rest-frame $(g-r)$ colour as a function of stellar mass. Central and satellite galaxies occupy complementary regions of the classical colour--mass relation, with satellites spanning a wider range of colours, particularly at low stellar masses, which may be connected with environmental quenching processes \citep[e.g.][]{2006MNRAS.366....2W,2013MNRAS.432..336W,2024MNRAS.527.5868M}. The colour-coding by the inner slope shows that red, low-mass satellite galaxies preferentially exhibit high values of $\gamma$, corresponding to steep, cusp-dominated inner density profiles. In many of these systems, the inferred slopes significantly exceed the canonical NFW value $\gamma=1$, indicating inner structures steeper than a standard NFW cusp. At higher stellar masses, both centrals and satellites occupy a narrower range of $\gamma$, with shallower inner profiles and a weaker dependence on colour.

The third column shows stellar mass as a function of the maximum circular velocity ($V_{\rm max}$). Central galaxies follow a relatively tight relation linking stellar content and internal kinematics \citep{2016MNRAS.460.3100C}, whereas satellites display a larger scatter, particularly at low $V_{\rm max}$. Across both populations, systems with lower $V_{\rm max}$ preferentially exhibit higher inner density slopes, indicating steeper, cusp-like inner profiles. This behaviour mirrors the trends observed for low-mass, red satellite galaxies discussed above, and suggests that galaxies with reduced central circular velocities tend to retain more concentrated inner DM distributions. Since $V_{\rm max}$ provides a robust tracer of the depth of the central gravitational potential \citep{2001MNRAS.321..559B,2011ApJ...740..102K}, lower values typically correspond to haloes that are both less massive and more susceptible to environmental processing. In particular, tidal stripping preferentially reduces $V_{\rm max}$ in satellite systems, while leaving the outer halo more strongly affected \citep{2004ApJ...608..663K,2008ApJ...672..904P}. This environmental dependence is most evident at low stellar masses. At higher stellar masses, however, the $M_\star$--$V_{\rm max}$ relation becomes tighter, and both central and satellite galaxies tend to exhibit lower $\gamma$ values, indicative of more core-like inner profiles and reduced differences between the two populations.

\begin{figure}[h!]
    \centering
    \includegraphics[width=0.9\linewidth]{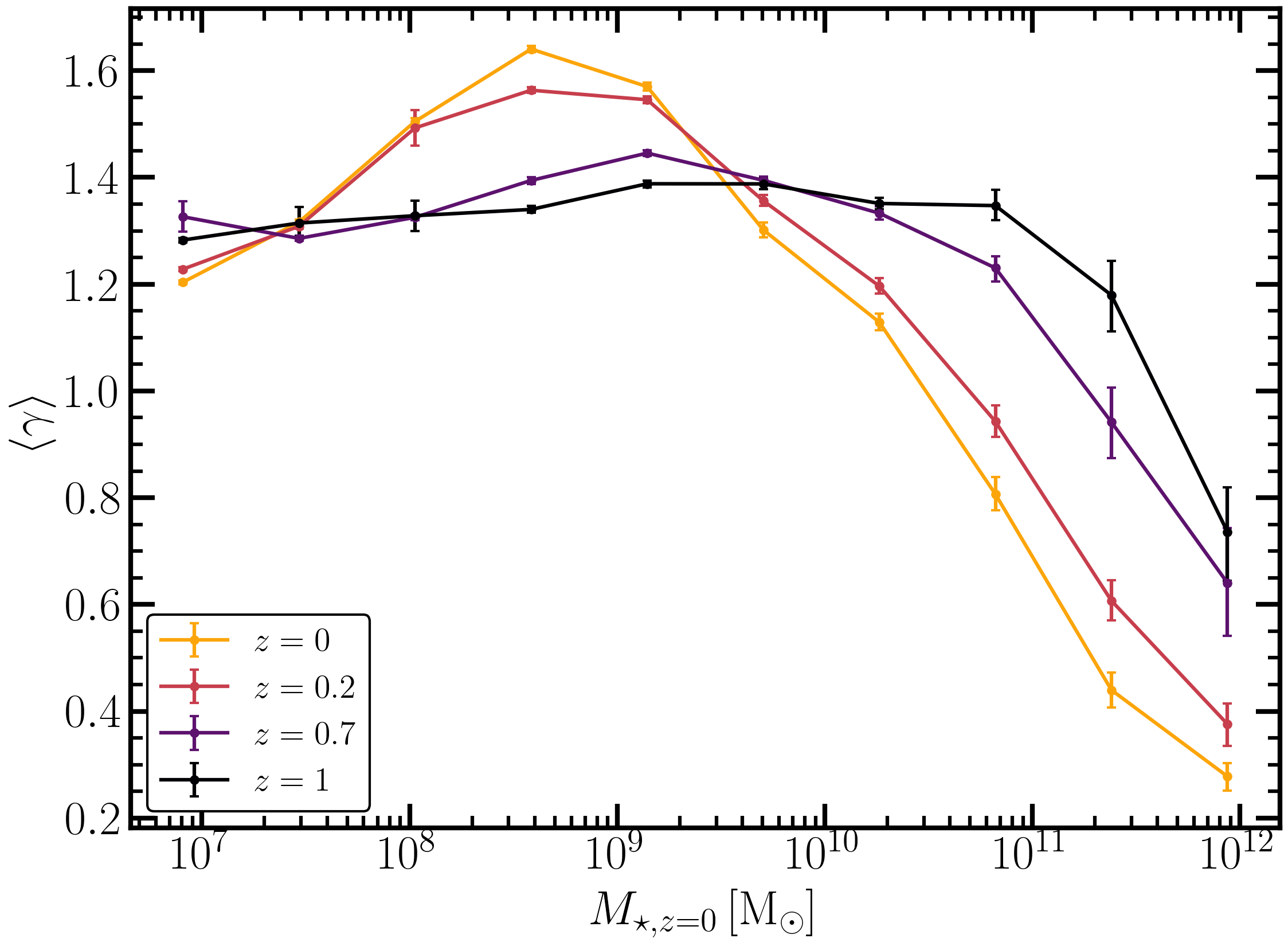}
    \caption{Mean inner slope ($\gamma$) as a function of the present-day stellar mass, $M_{\star,z=0}$, computed at redshifts $z=0$, $0.2$, $0.7$, and $1$. Each curve traces the main progenitors of the $z=0$ population. Error bars indicate bootstrap uncertainties.}
    \label{fig:gamma_stellar_mass_evol}
\end{figure}

\begin{figure*}[h!]
    \sidecaption
    \includegraphics[width=12cm]{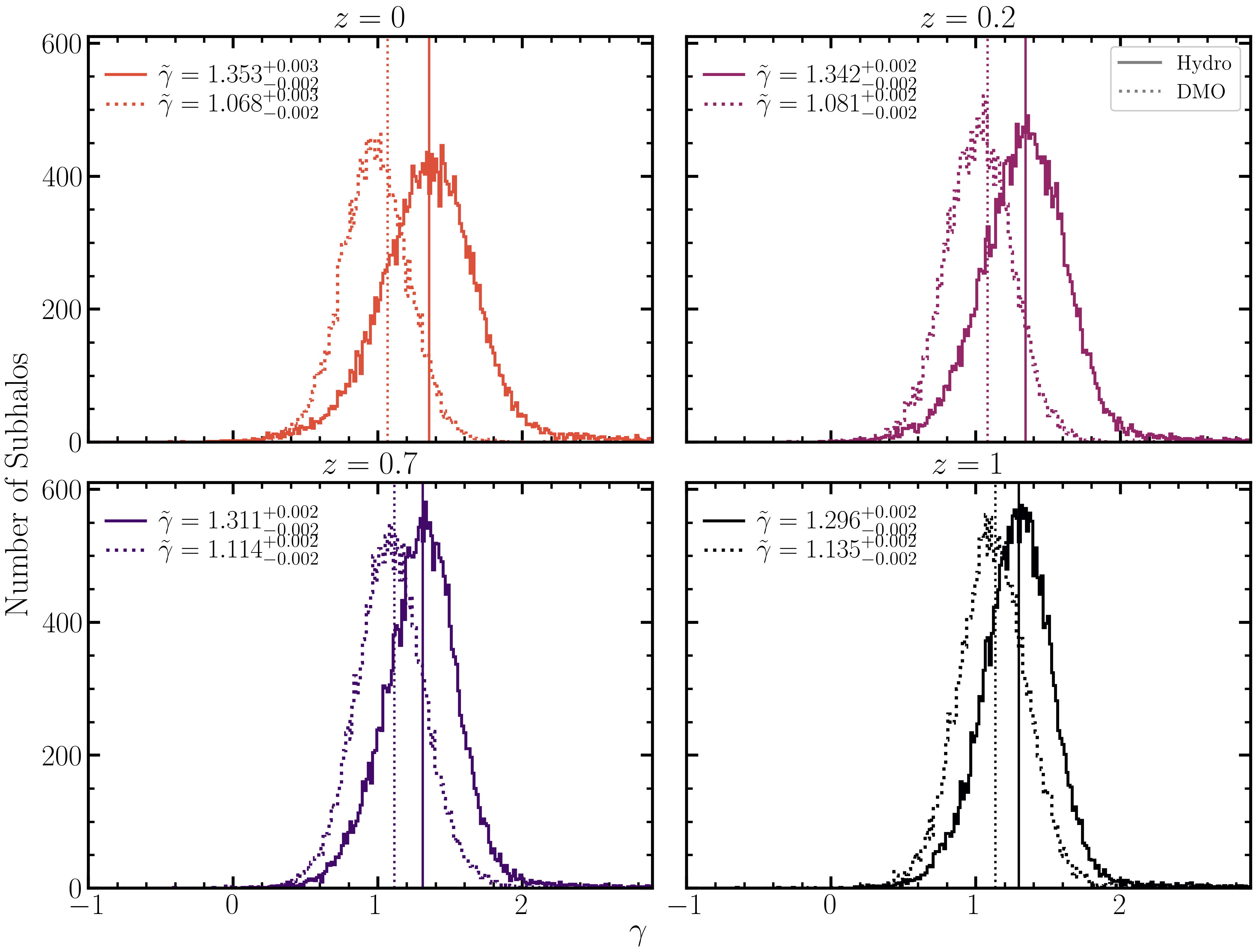}
    \caption{Distributions of the inner slope ($\gamma$) for the matched subhalo sample in TNG50 and its DMO counterpart at different redshifts. Vertical lines indicate the median values, with uncertainties given by the 16th--84th percentile confidence intervals from the bootstrap resampling.}
    \label{fig:slope_hydro_dmo}
\end{figure*}

The rightmost column displays $M_\star$ as a function of $M_{\rm host}$. Central galaxies trace the expected monotonic stellar-to-halo mass relation, reflecting the tight coupling between galaxy growth and halo assembly predicted by abundance-matching models \citep[e.g.][]{2013MNRAS.428.3121M,2013ApJ...770...57B}. Satellites, by contrast, span a much broader range of stellar masses at fixed host-halo mass, illustrating the diversity of accretion histories and environmental processing within massive haloes. Across this plane, systems with relatively low stellar masses embedded in massive host haloes tend to exhibit higher values of $\gamma$, corresponding to steeper inner density slopes. These objects largely coincide with the satellite population previously identified as having lower $V_{\rm max}$ and redder colours, indicating that the steepest inner profiles preferentially occur in satellites residing deep within massive environments.

\subsection{Redshift evolution} \label{subsec:evolution_in_redshift}

To investigate how the inner DM structure evolves over cosmic time, we traced its evolution over the last $\sim8$ Gyr by repeating the linear slope analysis from $z=1$ to $z=0$. Over this redshift range, the gravitational softening length for collisionless particles in TNG50 was fixed in physical units, which allowed us to use the same physical fitting interval at every epoch. We also verified that the distributions of $\chi_\nu^2$ remain low and comparable across redshifts, indicating a similar fit quality at all epochs.

Figure~\ref{fig:gamma_distribution_evol} shows the redshift evolution of the distribution of $\gamma$ from $z=1$ to $z=0$. The distribution gradually shifts towards higher values, with the median increasing from $\tilde{\gamma}=1.296^{+0.002}_{-0.002}$ at $z=1$ to $\tilde{\gamma}=1.353^{+0.003}_{-0.002}$ at $z=0$. Hereafter, the quoted uncertainties on median values correspond to the 16th--84th percentile intervals of their bootstrap distributions. Although monotonic, this evolution is modest, amounting to $\Delta\tilde{\gamma}=0.057$ over the redshift interval considered. At all redshifts, the median remains above the canonical NFW value and below the slopes characteristic of Moore-like profiles \citep{1999ApJ...524L..19M,1999MNRAS.310.1147M}. This trend indicates that, at the population level, the inner DM structure becomes steeper towards low redshifts. The broad distributions nevertheless reveal substantial system-to-system variation at each redshift.

A more detailed characterisation of this evolution is shown in Fig.~\ref{fig:gamma_stellar_mass_evol}, which displays the mean inner slope as a function of descendant stellar mass, $M_{\star,z=0}$. For each redshift, we identified the main progenitors of the $z=0$ population as described in Sect.~\ref{subsec:subhalo_sample}, and computed the mean $\gamma$ in bins of $M_{\star,z=0}$. The distributions within individual mass bins nevertheless exhibit substantial galaxy-to-galaxy scatter. This procedure links the evolution of the inner DM structure to the final stellar mass of each system, allowing us to trace how different galaxy populations build up their inner density profiles over cosmic time. In this framework, four distinct regimes of evolution can be identified when the figure is read vertically, i.e.\ at fixed $M_{\star,z=0}$, highlighting how the progenitors of present-day galaxies of different masses evolve in structurally diverse ways.

For the lowest-mass systems, around $M_{\star,z=0}\sim10^{7.5}\,{\rm M_\odot}$, the mean inner slope exhibits little evolution from $z=1$ to $z=0$, remaining close to $\langle\gamma\rangle\sim1.3$. At $M_{\star,z=0}\sim10^{8.6}\,{\rm M_\odot}$, the evolution becomes substantially stronger, with $\langle\gamma\rangle$ increasing from approximately $1.35$ at $z=1$ to $1.6$ at $z=0$. The largest change occurs between $z=0.7$ and $z=0.2$. Around $M_{\star,z=0}\sim10^{9.5}\,{\rm M_\odot}$, the slopes at all redshifts converge to $\langle\gamma\rangle\sim1.4$, showing only a weak dependence on redshift. At higher stellar masses, the evolutionary trend reverses: $\langle\gamma\rangle$ decreases progressively towards later times, with the separation between epochs becoming increasingly pronounced. Within this declining trend, the progenitors of galaxies with $M_{\star,z=0}\sim10^{10.8}\,{\rm M_\odot}$ and $\sim10^{11.4}\,{\rm M_\odot}$ exhibit mean inner slopes close to the canonical NFW value at $z=0.2$ and $z=0.7$, respectively. At the high-mass end, the evolution becomes particularly pronounced, leaving the most massive galaxies with substantially shallower inner profiles at $z=0$ than at $z=1$.

\subsection{Hydro versus DMO differences} \label{subsec:profiles_between_runs}

To quantify the impact of hydrodynamical physics on the inner DM structure, we compared TNG50 with its DMO counterpart, TNG50-Dark. The subhaloes of the two simulations were cross-matched using the Hydro--DMO associations described in Sect.~\ref{subsec:subhalo_sample}. This procedure allowed us to isolate the effects of baryons on the same underlying DM structures. Using this matched sample, we analysed both the full radial density profiles and the corresponding inner logarithmic slopes derived from linear fits over the resolved inner region. A representative comparison of matched subhaloes is shown in Fig.~\ref{fig:profiles_hydro_dmo}, illustrating the general agreement at large radii and the differences that arise in the inner regions.

The trend is illustrated in Fig.~\ref{fig:slope_hydro_dmo}, which compares the distributions of inner logarithmic slopes measured in TNG50 and TNG50-Dark for the full matched subhalo sample. At all redshifts considered, and consistently with the behaviour observed across the different mass bins, the hydrodynamical distributions are shifted towards higher values than their DMO counterparts. However, their median slopes evolve in opposite directions from $z=1$ to $z=0$: $\tilde{\gamma}_{\rm Hydro}$ increases, whereas $\tilde{\gamma}_{\rm DMO}$ decreases from $1.135^{+0.002}_{-0.002}$ to $1.068^{+0.003}_{-0.002}$. This DMO evolution is also modest, amounting to $\Delta\tilde{\gamma}_{\rm DMO}=-0.067$ over this interval (compared with $\Delta\tilde{\gamma}_{\rm Hydro}=0.057$, as reported in Sect.~\ref{subsec:evolution_in_redshift}). Consequently, the difference $\tilde{\gamma}_{\rm Hydro}-\tilde{\gamma}_{\rm DMO}$ increases from $\sim0.16$ at $z=1$ to $\sim0.29$ at $z=0$, showing that the separation between the Hydro and DMO distributions becomes progressively larger towards the present epoch.

\subsection{Robustness analysis} \label{subsec:diff_r_initial}

The measurement of the inner density slope depends sensitively on the radial interval over which the density profile is sampled, particularly in the innermost regions, where numerical resolution effects become significant. Throughout our analysis, we adopted $r_{\rm resol}=\epsilon_{\rm DM}$ as the reference resolution scale. For our fiducial measurement, we fitted the density profile over $[1,3]r_{\rm resol}$, which corresponds to the radial range described in Sect.~\ref{subsec:models}. We next examined how the inferred slopes depend on variations in this radial window.

\begin{figure}[h!]
    \centering
    \includegraphics[width=0.9\linewidth]{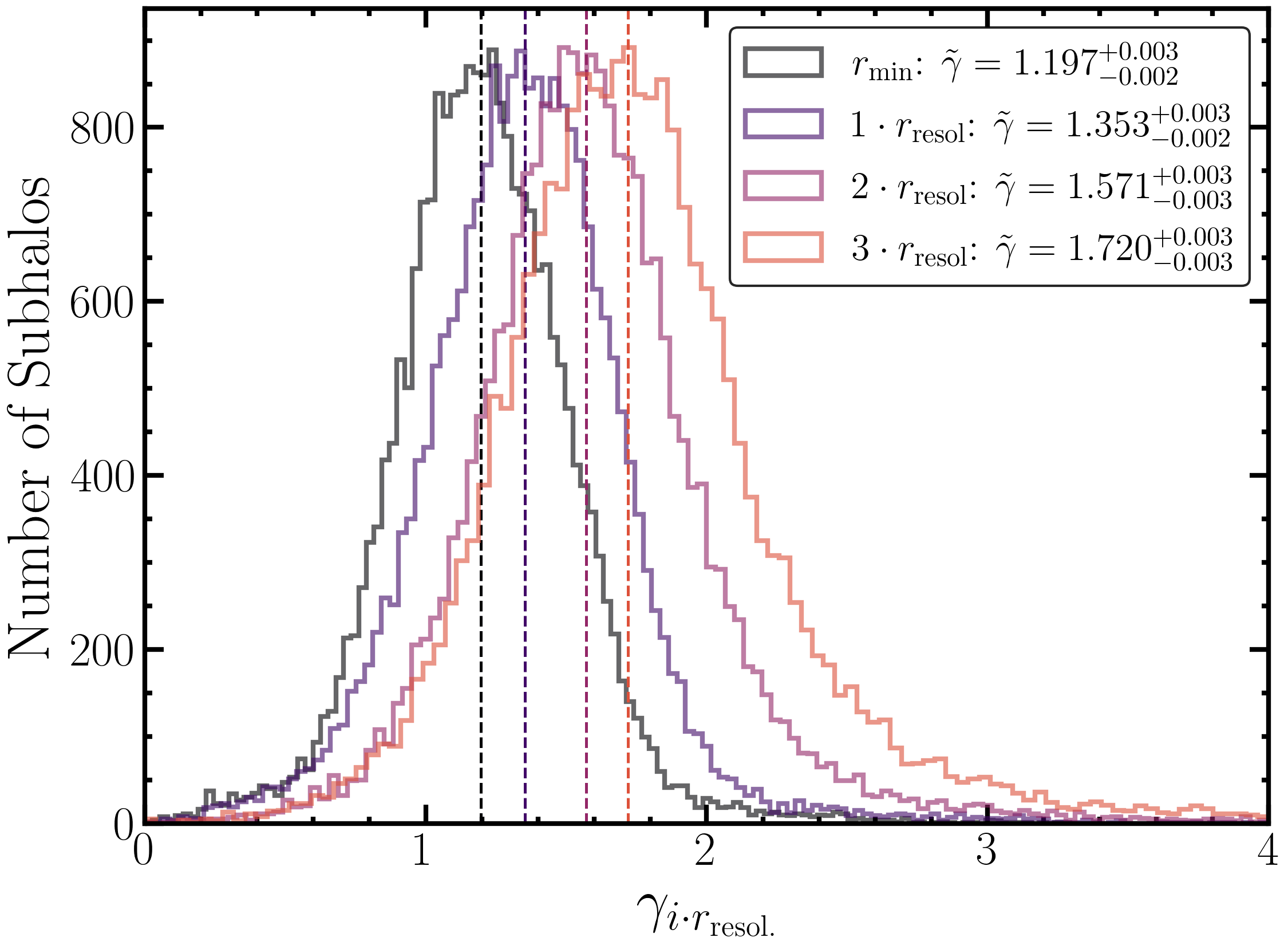}
    \caption{Distribution of the inner slope ($\gamma$) obtained from linear fits performed over different radial intervals: $[r_{\min},\,2\,r_{\rm resol}]$, $[1,3]\,r_{\rm resol}$ (fiducial), $[2,4]\,r_{\rm resol}$, and $[3,5]\,r_{\rm resol}$. Vertical dashed lines indicate the median values, with 16th--84th percentile uncertainties from bootstrap resampling. Shifting the fitting window towards larger radii leads to progressively steeper slopes, illustrating how excluding the central region suppresses shallow (core-like) profiles.}
    \label{fig2:r_res_tests}
\end{figure}

Density profiles were constructed starting from $r_{\min}=0.1\,\mathrm{kpc}$ (Sect.~\ref{subsec:subhalo_sample}), which lies below the softening length. In addition to the fiducial interval $[1,3]\,r_{\rm resol}$, we performed the linear fits over $[r_{\min},\,2\,r_{\rm resol}]$, $[2,4]\,r_{\rm resol}$, and $[3,5]\,r_{\rm resol}$. The resulting $\gamma$ distributions for the full galaxy sample are shown in Fig.~\ref{fig2:r_res_tests}. The inferred median slopes increase systematically as the fitting window is shifted outwards, from $\tilde{\gamma}=1.197^{+0.003}_{-0.002}$ for $[r_{\min},\,2\,r_{\rm resol}]$ to $\tilde{\gamma}=1.720^{+0.003}_{-0.003}$ for $[3,5]\,r_{\rm resol}$. This monotonic shift reflects the transition from mildly flattened inner regions to progressively cuspy profiles at larger radii. Consequently, adopting a more external fitting range biases $\gamma$ high by excluding the radii where shallow (core-like) behaviour is most prominent.

\begin{figure*}[h!]
    \centering
    \includegraphics[width=0.9\linewidth]{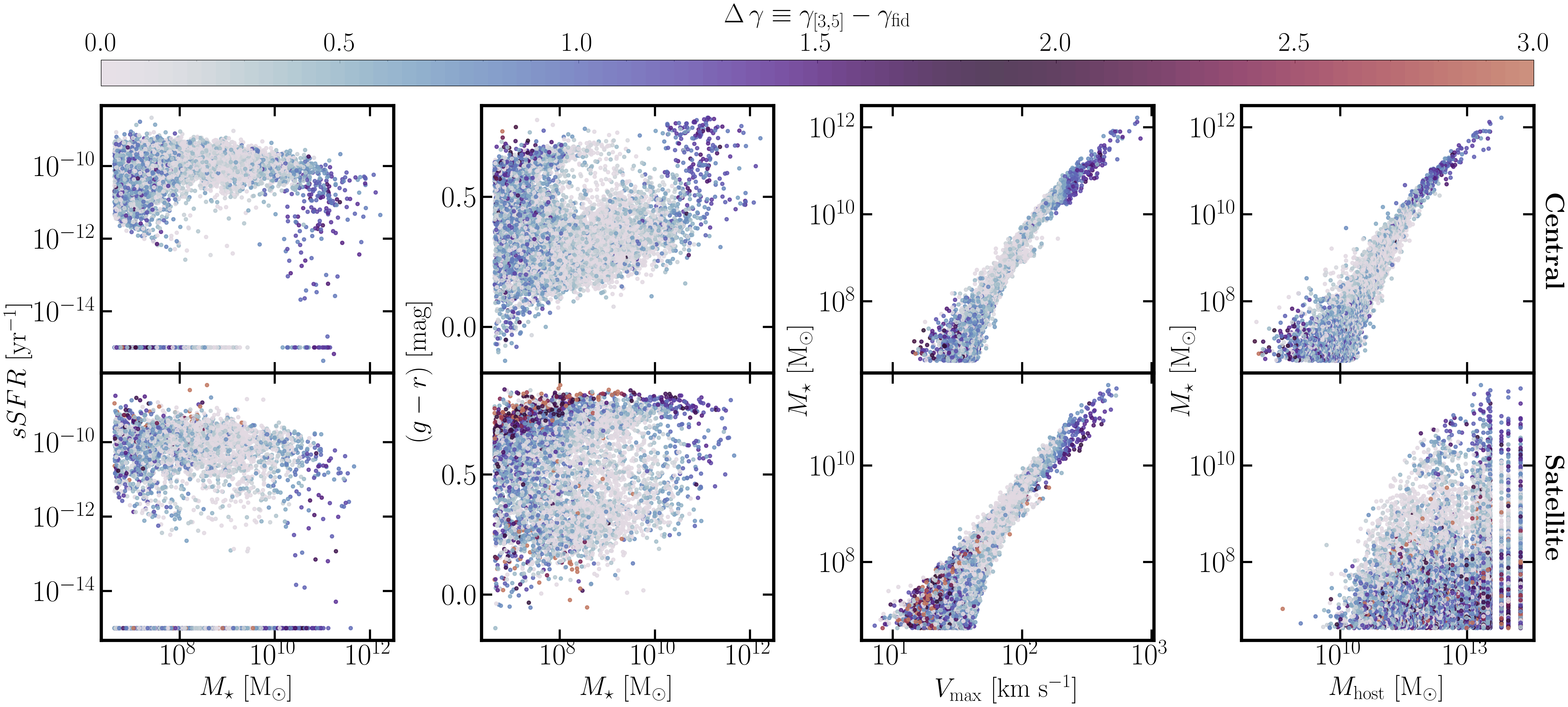}
    \caption{Difference in the inner density slope between a more external fitting window and the fiducial fit, colour-coded as $\Delta \gamma \equiv \gamma_{[3,5]} - \gamma_{\rm fid}$, where $\gamma_{\rm fid}\equiv\gamma_{[1,3]\,r_{\rm resol}}$ and $\gamma_{[3,5]}\equiv\gamma_{[3,5]\,r_{\rm resol}}$.\ We can see how the inferred inner slope changes when adopting a more external fitting window across galaxy properties. From left to right, columns show: the sSFR--$M_\star$ plane; the rest-frame $(g-r)$--$M_\star$ relation; the $M_\star$--$V_{\rm max}$ relation; and the $M_\star$--$M_{\rm host}$ relation. Top and bottom rows correspond to central and satellite galaxies, respectively.}
    \label{fig:delta_gamma_color}
\end{figure*}

Building on this result, Fig.~\ref{fig:delta_gamma_color} shows the difference between the slopes measured within the fiducial interval $[1,3]\,r_{\rm resol}$ and those obtained over $[3,5]\,r_{\rm resol}$, colour-coded as $\Delta \gamma \equiv \gamma_{[3,5]} - \gamma_{\rm fid}$, across the same galaxy scaling relations presented in Fig.~\ref{fig:2x4_figure}, separately for central (top panels) and satellite (bottom panels) galaxies. The largest variations are found at both the low- and high-stellar-mass ends, whereas galaxies in the intermediate mass range ($M_\star \sim 10^{8}$--$10^{10}\,{\rm M_\odot}$) show much smaller differences. In the high-mass regime, systems that appear shallow at the fiducial radius tend to exhibit a pronounced steepening when the innermost region is excluded. Conversely, in the low-mass regime, systems with the steepest inner slopes typically become even steeper when the inner region is excluded, with no indication of an opposite trend that would make the result dependent on the adopted fitting range. This is particularly important, as it demonstrates that the population identified as having the steepest slopes is robust against variations in the fitting procedure.

This behaviour is consistently observed across the sSFR--$M_\star$, $(g-r)$--$M_\star$, $M_\star$--$V_{\rm max}$, and $M_\star$--$M_{\rm host}$ relations, and in both central and satellite populations, indicating a mass-dependent impact of the adopted fitting window. While negative values of $\Delta \gamma$ are in principle possible, the distribution is predominantly shifted towards positive values (see Fig.~\ref{fig2:r_res_tests}). Overall, excluding the innermost resolved region yields higher inferred inner slopes (i.e. predominantly $\Delta \gamma > 0$), supporting the use of the fiducial interval. This mainly affects high-mass systems, increasing their slopes and attenuating the overall mass dependence, while low-mass galaxies remain consistently steeper.

\section{Discussion}
\label{sec:discussion}

\subsection{Cosmic evolution of inner DM profiles}

At the population level, TNG50 shows a modest but monotonic evolution in its inner DM structure, with the median slope increasing from $\tilde{\gamma}=1.296$ at $z=1$ to $\tilde{\gamma}=1.353$ at $z=0$, while remaining steeper than NFW throughout. The progenitor tracks nevertheless reveal a mass-dependent evolution. At earlier epochs, repeated feedback-driven outflows can induce potential fluctuations that flatten inner DM profiles \citep{2012MNRAS.421.3464P,2012MNRAS.422.1231G}. The progenitors of galaxies with $M_{\star,z=0}\sim10^{10.8}$ and $10^{11.4}\,{\rm M_\odot}$ reach $\gamma\sim1$ at $z=0.2$ and $z=0.7$, respectively. These relatively shallow values are qualitatively compatible with feedback-driven flattening.

Two recent studies provide complementary comparisons for distinct galaxy populations. The TNG100 analysis of \citet{2025MNRAS.541....1S}, based on the elliptical-galaxy sample of \citet{2019MNRAS.490.5722W}, finds that galaxies with $10^{10.7}\leq M_{\star,z=0}\leq10^{11.9}\,{\rm M_\odot}$ develop steeper profiles towards $z=0$, whereas our unrestricted TNG50 sample becomes shallower over the same mass range. Conversely, \citet{2026A&A...708A.112C} find $\gamma<0.5$ for $66\%$ of 127 star-forming galaxies at $0.3<z<1.5$, while their mass-matched TNG50 star-forming sample is almost exclusively cuspy. These contrasting results suggest that galaxy selection may contribute to the reported differences. Separating TNG50 galaxies by morphology and star-formation activity would allow this possibility to be tested in future work.

At $z=0$, \citet{2014MNRAS.437..415D}, \citet{2016MNRAS.456.3542T}, and \citet{2020MNRAS.497.2393L} find a non-monotonic dependence of the inner slope on $M_\star/M_{\rm halo}$, with maximum flattening around $M_\star/M_{\rm halo}\sim(4$--$6)\times10^{-3}$. Figure~\ref{gamma_vs_ratio_mass} shows that TNG50 central galaxies primarily follow the subsequent steepening branch, although shifted towards lower stellar-to-halo mass ratios. Within the contour enclosing $50\%$ of the distribution, the shallowest profiles occur at $M_\star/M_{\rm halo}\simeq8\times10^{-4}$. The TNG50 values of $-\gamma$ are also generally more negative, indicating steeper profiles, although the broad distribution overlaps part of the range reported in these studies. These offsets may partly arise from differences in radial intervals, slope definitions, numerical methods, resolution, and feedback prescriptions among the simulations.

\subsection{Effect of baryons}

The comparison between TNG50 and its DMO counterpart indicates that baryons systematically affect the inner DM structure at all redshifts explored. When measured over identical radial intervals, hydrodynamical galaxies exhibit steeper inner slopes than their DMO counterparts.

This offset is qualitatively consistent with the net effect of baryonic condensation, which can enhance the central DM density and produce more centrally concentrated profiles relative to their DMO counterparts \citep[e.g.][]{2009MNRAS.395L..57P, 2010MNRAS.407..435A, 2019A&A...622A.197A, 2026ApJ..1003...12L}. The strength and even the sign of this response, however, depend on the adopted feedback model and halo assembly history \citep{2009MNRAS.395L..57P, 2010MNRAS.405.2161D}. This interpretation is further supported by \citet{2026ApJ..1002....8G}, who analysed matched hydro--DMO realisations from the dark matter and astrophysics with machine learning and simulations (DREAMS) suite \citep{2025arXiv251200148R} and find that, at $z=0$, the baryonic response of Milky Way-mass haloes is generally consistent with adiabatic contraction across the explored variations in the TNG model and cosmology. Although the baryonic response can vary among individual haloes, the systematic shift of the hydrodynamical distributions towards higher $\gamma$ at all redshifts indicates a baryonic imprint on the inner halo structure at the population level.

A complementary perspective is provided by high-resolution zoom-in simulations. For example, \citet{2026arXiv260113765K} analysed the evolution of the inner DM cusp in dwarf galaxies and find that baryonic physics can steepen the inner profile in systems that assemble early and develop deep central potentials. Despite differences in numerical approach and resolution, our sample of matched galaxies exhibits qualitatively similar behaviour: hydrodynamical galaxies tend to display steeper inner slopes than their DMO counterparts across all redshifts. Together, these results suggest that baryon-induced modifications of the inner DM profile are a robust outcome across different numerical approaches.

\subsection{Physical drivers of inner DM structure}

Stellar mass is best regarded as an integrated proxy, rather than a unique driver, of the baryonic response of the inner halo. It traces both the depth of the central baryonic potential and a cumulative assembly history shaped by star formation within the galaxy and stellar accretion through mergers \citep[e.g.][]{2014MNRAS.444.1453D,2016MNRAS.458.2371R}. As discussed above, the resulting DM response reflects the opposing effects of feedback-driven potential fluctuations and the contraction induced by centrally concentrated baryons \citep[e.g.][]{2013MNRAS.432.1947M,2018MNRAS.480..800H}. The relative importance of these processes provides a qualitative interpretation of the non-monotonic dependence of $\gamma$ on $M_\star/M_{\rm halo}$ and cautions against interpreting trends with individual galaxy properties as arising from a single mechanism.

Within the hydrodynamical sample, the dependence of $\gamma$ on stellar mass differs between central and satellite galaxies, indicating that the response of the inner halo varies across the galaxy population. At high stellar masses, however, both populations exhibit similarly shallow slopes. Such profiles may reflect baryonic processes that redistribute mass in the central regions and thereby modify the gravitational potential \citep[e.g.][]{2017MNRAS.472.2153P,2020MNRAS.491.5188W}. At lower stellar masses, satellites exhibit a broader diversity of slopes, with systematic variations related to colour, $V_{\rm max}$, and host-halo mass.

In particular, the steepest slopes are preferentially found among low-mass satellites that are red, have lower $V_{\rm max}$, and reside in more massive host haloes. Their red colours are consistent with a quenched population, while their association with massive hosts suggests a possible role for environmental mechanisms such as ram-pressure stripping or strangulation. By removing the gas reservoir or preventing its replenishment, these processes may suppress sustained baryonic cycling and limit feedback-driven potential fluctuations \citep[e.g.][]{1972ApJ...176....1G,2008MNRAS.383..593M,2013MNRAS.432..336W}. In this scenario, reduced feedback activity would limit core formation and favour the persistence of steeper inner profiles. However, a dedicated environmental analysis would be required to quantify the relative importance of these mechanisms.

\subsection{Comparison with observational constraints} \label{disc:observations}

In the low-mass regime, observational estimates for dwarf galaxies generally favour shallower inner profiles than those found in TNG50. Using H\,{\sc i} rotation curves probing the inner ${\sim}1\,\mathrm{kpc}$ of gas-rich dwarfs from the Local Irregulars That Trace Luminosity Extremes, The H\,{\sc i} Nearby Galaxy Survey (LITTLE THINGS; \citealt{2012AJ....144..134H}), \citet{2015AJ....149..180O} measured a mean slope of $\gamma=0.32\pm0.24$ and pseudo-isothermal core radii down to ${\sim}0.15\,\mathrm{kpc}$. If similarly compact cores exist in TNG50, they could lie below our fiducial range, causing the ILF to sample the transition towards the outer profile and contributing to the larger inferred slopes. A closer comparison in terms of both stellar mass and radial scale is provided by \citet{2022MNRAS.512.1012C}, who measured inner DM slopes over $0.3$--$0.8\,\mathrm{kpc}$ for six nearby spiral dwarf galaxies spanning $M_\star=10^{9.3}$--$10^{9.7}\,{\rm M_\odot}$. Their mean slope of $\gamma=0.6$, with a galaxy-to-galaxy scatter of $0.1$, remains below the corresponding TNG50 values, although the simulated mean varies considerably across the same stellar-mass interval.

Beyond this general tendency towards shallow profiles, some studies reveal considerable galaxy-to-galaxy diversity. \citet{2019MNRAS.484.1401R} analysed eight LITTLE THINGS dwarf irregulars and eight classical Milky Way dwarf spheroidals, finding higher central densities and steeper local profiles in systems quenched more than $6\,\mathrm{Gyr}$ ago. Their local slopes, however, were evaluated at $150\,\mathrm{pc}$, below our fiducial interval, and remain sensitive to the adopted priors. Using the same eight classical dwarf spheroidals, \citet{2020ApJ...904...45H} likewise find a broad range of inner slopes, with most galaxies favouring cusped profiles. Recent analyses of Draco and Sculptor further illustrate this diversity, yielding asymptotic slopes of $\gamma=0.71^{+0.34}_{-0.35}$ and $\gamma=0.39^{+0.23}_{-0.26}$, respectively \citep{2025ApJ...993..249Y,2025A&A...699A.347A}. Although both galaxies lie below our stellar-mass limit, these results suggest that the inferred diversity may reflect differences in both star formation histories and dynamical modelling.

The comparison also depends on the radial scale probed. For Sculptor, \citet{2026arXiv260708818S} find that the slope increases from $\gamma=0.44^{+0.14}_{-0.13}$ at $100\,\mathrm{pc}$ to $\gamma=0.75^{+0.11}_{-0.10}$ at $500\,\mathrm{pc}$. This steepening towards larger radii is qualitatively consistent with the steeper slopes obtained when shifting our ILF fitting interval outwards. A further complication at these scales is the presence of unresolved central baryonic clumps within $r\lesssim1\,\mathrm{kpc}$ in low-mass isolated TNG50 galaxies \citep{2025A&A...699A..12C}. Although their impact on the DM density profiles has not been quantified, their overlap with our fiducial interval may introduce additional uncertainty in the inferred slopes. Overall, TNG50 also exhibits a broad range of slopes at low masses, although its highest values of $\gamma$ remain difficult to reconcile with the observational estimates considered here.

At larger scales, constraints on the inner DM distribution of galaxy clusters can be obtained by combining gravitational lensing with the stellar kinematics of their brightest cluster galaxies. Using this approach, \citet{2013ApJ...765...25N} measured a mean asymptotic gNFW slope of $\gamma=0.50\pm0.10\,(\mathrm{random})\,{}^{+0.14}_{-0.13}\,(\mathrm{systematic})$ for seven relaxed clusters, while \citet{2025MNRAS.541.2341C} obtain a mean slope of $\gamma\simeq0.66$ over ${\sim}5$--$50\,\mathrm{kpc}$ for six clusters hosting radial arcs. Although these constraints are derived from data that probe larger physical radii than our ILF, standard monotonic halo profiles are expected to remain shallow or flatten further towards their asymptotic inner slopes. Their reported values of $\gamma$ therefore provide a useful observational reference for the high-mass TNG50 profiles.

\section{Conclusions} \label{sec:conclusions}

In this work we investigated the inner DM density structure of TNG50 galaxies using the ILF, which measures the inner slope through a linear fit in logarithmic space while explicitly accounting for the adopted radial fitting range. This approach provides a consistent description of the central DM distribution across galaxy populations, redshifts, and simulation runs. Our main conclusions can be summarised as follows:

\begin{enumerate}

\item The inner DM density structure of TNG50 galaxies shows a modest but systematic evolution towards steeper profiles at lower redshifts, while the DMO run follows the opposite trend. Nevertheless, hydrodynamical galaxies remain systematically steeper than their DMO counterparts at all redshifts, suggesting that baryonic processes increase the central DM concentration, consistent with quasi-adiabatic contraction.

\item The inner slope ($\gamma$) shows a clear dependence on stellar mass and galaxy properties, with high-mass systems preferentially exhibiting shallow or weakly cusped profiles, with no significant differences between central and satellite galaxies, indicating that baryonic processes are likely to dominate over environmental effects in this regime. At lower masses, however, $\gamma$ displays a much broader diversity that correlates with galaxy properties and may also reflect environmental effects.

\item At fixed stellar mass, redder satellites in the low-mass regime tend to exhibit steeper inner slopes, with the highest $\gamma$ values preferentially associated with lower $V_{\rm max}$ and more massive host haloes. If this steepness is not primarily driven by unresolved inner flattening, it may instead reflect reduced feedback-driven potential fluctuations in quenched systems.

\item Comparisons with observations reveal a mass-dependent picture. At low masses, observational estimates generally favour shallower inner slopes than those found in TNG50, although this discrepancy could partly reflect compact cores lying below our fiducial fitting interval. At the high-mass end, the flattening of the inner profiles is qualitatively consistent with cluster-scale measurements, since standard monotonic halo profiles are expected to remain shallow or flatten further towards smaller radii.

\item The inferred inner slopes depend on the adopted radial fitting range, with more external intervals generally yielding steeper profiles. Intermediate-mass galaxies are the least affected, whereas the reddest low-mass satellites remain among the most cuspy systems and become even steeper when the innermost region is excluded. This indicates that, although the innermost flattening of low-mass systems may occur below the resolved scales probed here, the fiducial interval provides the most suitable reference for identifying the shallowest core-like structure accessible at the resolution limit.

\end{enumerate}

\begin{acknowledgements}
We thank the anonymous referee for valuable comments and suggestions that improved the manuscript. VH and ADMD acknowledge support from the Universidad Técnica Federico Santa María through the Proyecto Interno Regular \texttt{PI\_LIR\_25\_04}. MCA acknowledges financial support from Fondecyt Iniciación number 11240540. ADMD, MCA, and AK acknowledge partial support from the ALMA fund with code 31220021. MCA and AK acknowledge partial support from ANID BASAL project FB210003.
\end{acknowledgements}

\bibliographystyle{aa} 
\bibliography{references}

\begin{appendix}

\onecolumn

\section{Illustrative density profiles across runs}

\begin{figure*}[ht!]
    \sidecaption
    \includegraphics[width=12cm]{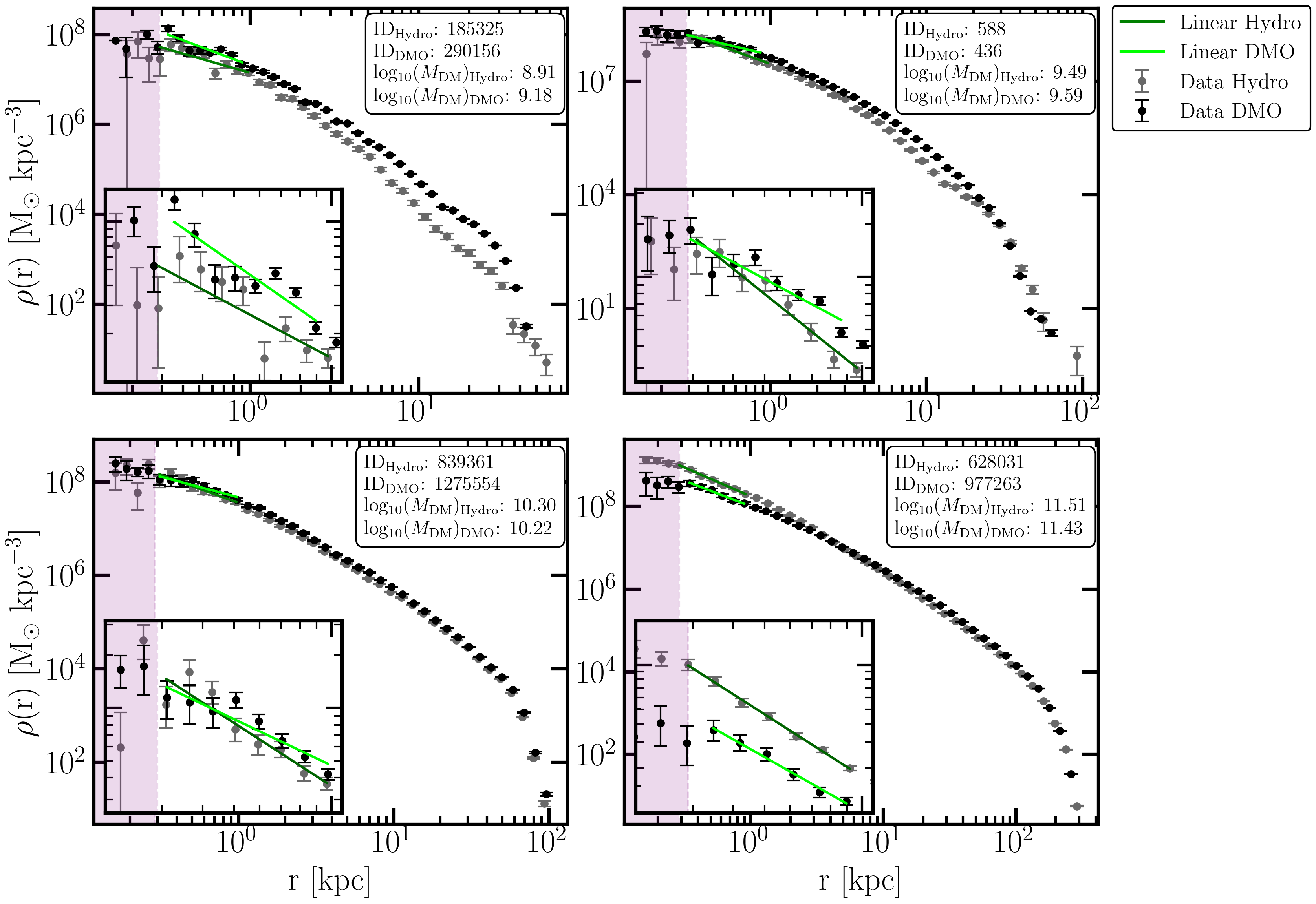}
    \caption{Hydrodynamical versus DMO density profiles for the systems shown in Fig.~\ref{fig1:fits_all_prof}. Each panel shows the full radial profile along with a zoomed-in view of the inner region and the corresponding linear fits. The overall shape of the profiles agrees between runs, while systematic differences arise in the inner regions, where the hydrodynamical version is typically steeper.}
    \label{fig:profiles_hydro_dmo}
\end{figure*}

Figure~\ref{fig:profiles_hydro_dmo} presents representative density profiles of the same matched subhaloes shown in Fig.~\ref{fig1:fits_all_prof}, now displayed separately for the hydrodynamical and DMO runs. The examples span more than two orders of magnitude in DM mass and are intended to illustrate the comparison procedure between the two simulations. Overall, the hydrodynamical and DMO profiles show similar shapes, indicating that the global halo structure is largely preserved between the runs. Differences emerge in the inner regions, where, in these representative cases, the hydrodynamical profiles are consistent with steeper inner logarithmic slopes than their DMO counterparts. The ILF provides a quantitative framework to characterise such differences in the inner profiles. We stress that this figure is illustrative and does not imply that hydrodynamical haloes are uniformly steeper than DMO haloes on an object-by-object basis; rather, it motivates the statistical comparison of inner slopes across the full matched sample.

\FloatBarrier

\section{Inner DM slope and stellar-to-halo mass ratio}
\label{app:complementary_trends}

\begin{figure}[ht!]
    \sidecaption
    \includegraphics[width=8.5cm]{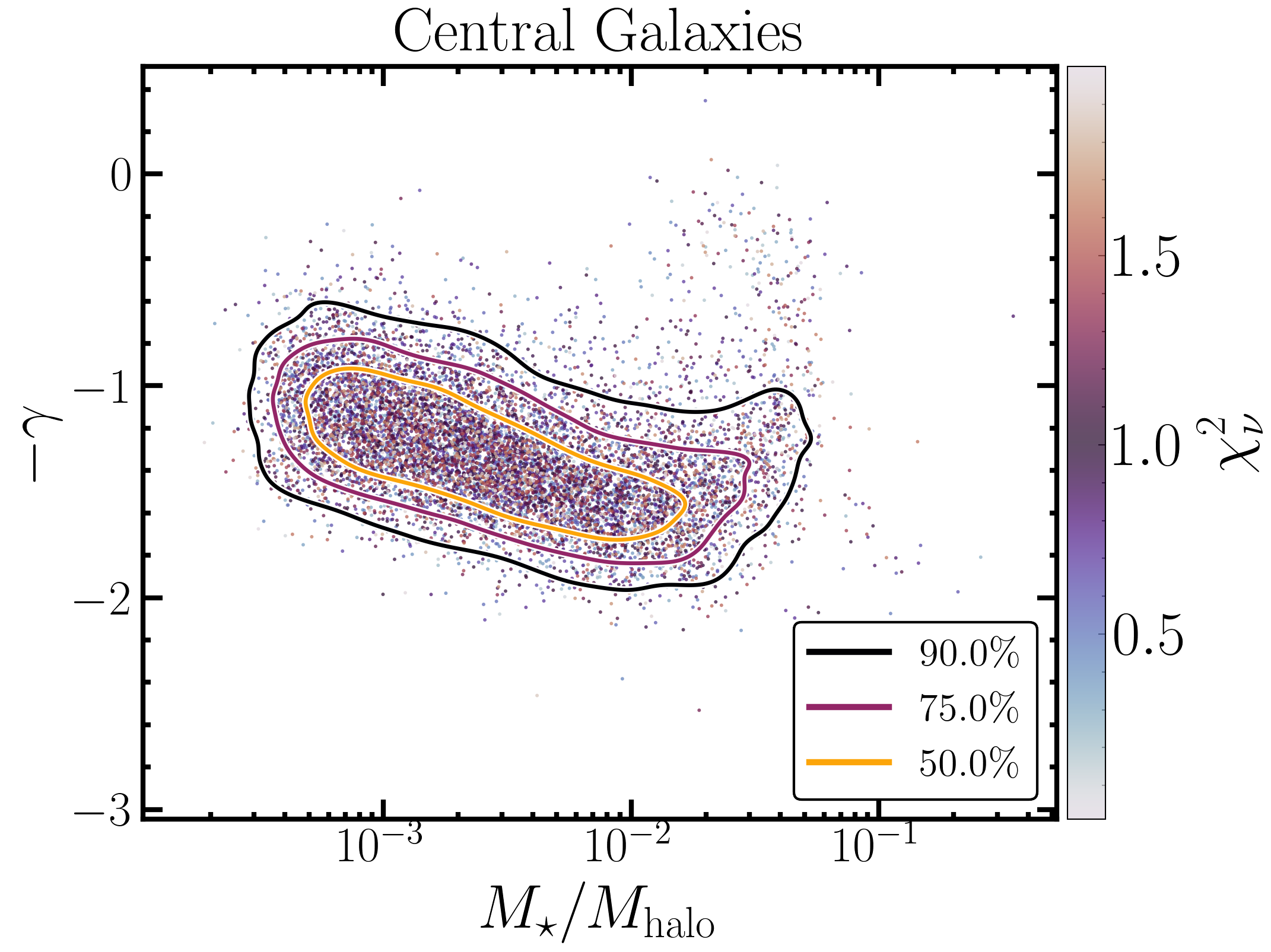}
    \caption{Inner logarithmic DM density slope, $-\gamma$, as a function of the stellar-to-halo mass ratio for central galaxies at $z=0$. In this work, the slopes were obtained from the fiducial linear fit over $[1,3] \, r_{\rm resol}$. For comparison, \citet{2014MNRAS.437..415D}, \citet{2016MNRAS.456.3542T}, and \citet{2020MNRAS.497.2393L} denoted the logarithmic slope as $\alpha=-\gamma$ and evaluated it over the scaled radial interval $0.01<r/R_{\rm vir}<0.02$. Individual galaxies are coloured according to $\chi_\nu^2$, and the contours enclose $50$, $75$, and $90\%$ of the distribution. The relation exhibits substantial scatter, with the highest-density region shifting towards steeper slopes as $M_\star/M_{\rm halo}$ increases.}
    \label{gamma_vs_ratio_mass}
\end{figure}

Figure~\ref{gamma_vs_ratio_mass} shows that, over the range populated by most TNG50 central galaxies, the inner profiles tend to become steeper with increasing $M_\star/M_{\rm halo}$. This behaviour resembles the steepening branch of the non-monotonic relations reported by \citet{2014MNRAS.437..415D}, \citet{2016MNRAS.456.3542T}, and \citet{2020MNRAS.497.2393L}, although it begins at a lower stellar-to-halo mass ratio. The shallowest part of the region enclosing $50\%$ of the TNG50 galaxies occurs near $M_\star/M_{\rm halo}\simeq8\times10^{-4}$, compared with the maximum flattening at $\sim(4$--$6)\times10^{-3}$ in those studies. The TNG50 distribution also extends towards more negative values of $\alpha$, indicating generally steeper profiles, although its broad scatter overlaps part of the previously reported range. These differences may partly reflect the adopted radial intervals and slope definitions, as well as differences in numerical methods, resolution, and feedback prescriptions.

\FloatBarrier

\end{appendix}

\end{document}